\documentclass[12pt]{iopart}
\usepackage[utf8]{inputenc}
\usepackage[russian]{babel}
\usepackage{amscd}
\usepackage{bbm}
\usepackage{bm}
\usepackage{xy}
\usepackage{amssymb}
\usepackage{graphicx}
\usepackage{theorem}
\usepackage{makeidx}
\usepackage{epic}
\usepackage{epsfig}
\usepackage{times}
\usepackage{verbatim}
\usepackage[T1]{fontenc}
\usepackage{color}

\usepackage[X2,T2A]{fontenc}
     \DeclareTextSymbolDefault{\CYRYAT}{X2}
     \DeclareTextSymbolDefault{\CYRFITA}{X2}
     \DeclareTextSymbolDefault{\CYRIZH}{X2}
     \DeclareTextSymbolDefault{\cyryat}{X2}
     \DeclareTextSymbolDefault{\cyrfita}{X2}
     \DeclareTextSymbolDefault{\cyrizh}{X2}

\newtheorem{Theorem}{Theorem}[section]
\newtheorem{Prop}[Theorem]{Предложение}

\begin{document}

\title[Математические аспекты задачи коррекции оптической поверхности]{Математические аспекты задачи коррекции оптической поверхности травлением ионными пучками}
\author{Вальтер Паульс$^{1}$}

\address{
$^1$Институт физики микроструктур РАН, 603087 д. Афонино, Кстовский р-н, Нижегородская обл., Россия}
\ead{walter.pauls@gmail.com}
\date{}

\begin{abstract}

Основываясь на задаче расчета карты травления при обработке поверхности с  помощью ионного пучка (Ion Beam Figuring), мы предлагаем математический аппарат для решения 
определённого типа обратных задач, которые возникают в различных областях прикладной математики и физики.
С точки зрения функционального анализа мы имеем дело с линейным операторным уравнением, которое может быть решено с использованием псевдообратного оператора. 
При этом оказывается, что оператор, связанный с этим уравнением и действующий из пространства карт травления в пространство результатов измерений, близок к конечномерному,. 
Для основного интересующего нас случая мы описываем поведение сингулярных векторов внутри области, на которой заданы данные измерений; эти векторы оказываются близки к функциям $e^{ikx}$. Приближенно, эти сингулярные векторы похожи на собственные функции оператора Гамильтона бесконечно глубокой квантовой ямы. Альтернативная постановка задачи, более близкая к практическим расчётам, использует гильбертовы пространства с воспроизводящими ядрами (RKHS) и радиальные базисные функции, которые применяются для точечной аппроксимации результатов измерений и последующего расчета  карты травления. В зависимости от особенностей конкретной задачи может быть использован любой из этих подходов или их сочетание.

\end{abstract}




\section{Введение}

В последние годы обработка поверхностей с помощью облучения пучками частиц стала стандартным инструментом в производстве оптических поверхностей\footnote{Данная работа является расширенным русским переводом статьи \cite{JPhysA}. Предварительные результаты были опубликованы в патенте \cite{Patent}.}. Особенно широко распространено ионно-пучковое травление (IBF) \cite{WM87, WRM88,1}, которое использует удаление материала подложки (распыление) при ионной бомбардировке: пучок ускоренных ионов, сфокусированный на обрабатываемой поверхности, вытравливает материал подложки в количестве, пропорциональном времени экспозиции пучка в указанной точке. Размещая сфокусированный ионный пучок над различными участками поверхности и соответствующим образом выбирая время его экспозиции, можно достичь требуемого изменения профиля поверхности.
Конкретный вид зависимости количества удаляемого материала подложки от локального времени экспозиции варьируется для разных установок ионно-пучкового травления в соответствии с их конструктивными особенностями, поскольку существует широкое разнообразие инженерных реализаций этой технологии — от лабораторных установок \cite{2,6,5} до коммерческого оборудования для промышленного применения. Следует отметить, что аспект, наиболее существенный для нашего обсуждения, — это механизм сканирования сфокусированного пучка по обрабатываемой поверхности. Как будет показано далее, выбор между непрерывным и дискретным перемещением пучка определяет возможность использования различных математических моделей процесса травления. Определённую роль также играют степени свободы перемещения ионного источника, используемая система координат и геометрия обрабатываемых поверхностей, однако рассмотрение этих вопросов выходит за рамки настоящего исследования.

Предположим, что профиль поверхности, подлежащий коррекции, задан в виде функции $h(x,y)$ от координат $(x,y)$ на поверхности. Также предположим, что мы знаем из измерений локальную скорость распыления ионного пучка, расположенного в начале координат, в виде функции $f(x,y)$ от координат $(x,y)$. Другими словами, $f(x,y)$ описывает функциональную форму полости, создаваемой в подложке за единицу времени под воздействием ионного пучка. Она зависит от размера пятна сфокусированного ионного пучка, тока ионного пучка, ускоряющего напряжения и различных других параметров. В дальнейшем мы будем называть $f(x,y)$ \textbf{(рабочей) функцией пучка} (англ. tool influence function, нем. Werkzeugfunktion).

Математическая задача в общих чертах состоит в вычислении карты времени экспозиции ионного пучка по поверхности (карты травления), которая приведёт к изменению профиля, максимально близкому к заданному, т.е. к $h(x,y)$. В простейшем случае, который мы в основном будем рассматривать в этой статье, мы предполагаем, что время экспозиции ионного пучка является непрерывной функцией $t(x,y)$ координат поверхности $(x,y)$. Изменение профиля в результате ионно-лучевого травления со временем экспозиции, распределённым в соответствии с $t(x,y)$, задаётся свёрткой карты травления 
$t(x,y)$ с функцией пучка $f(x,y)$:
\begin{displaymath}
(t \ast f)(x,y) = \int  f (x - x^{\prime } , y - y^{\prime } ) t (x^{\prime } , y^{\prime }) \, dx^{\prime } dy^{\prime } .
\end{displaymath}
Здесь мы неявно предположили, что обрабатываемая поверхность близка к плоской. Отметим, что также можно задать время экспозиции на дискретной сетке $\Delta \mathbb{Z}^2 $, где $\Delta $ — постоянная решётки, что тогда приводит к выражению суммами вместо интегралов. В физическом и инженерном сообществе установились два типа подходов к вычислению карты травления для заданных профилей коррекции поверхности: (i) решения с использованием преобразования Фурье \cite{WM87,7,8} и (ii) решения, основанные на методах линейной алгебры \cite{9,10}.

При определении карты травления для заданной коррекции профиля поверхности возникает несколько проблем. Во-первых, хорошо известно, что «деконволюция» подобного рода является некорректно поставленной задачей в смысле Адамара. Во-вторых, в общем случае карта коррекции поверхности $h(x,y)$ задана только на ограниченной области, которую мы называем областью коррекции или
областью измерений, вне которой она не определена. Однако время экспозиции должно быть определено на области, которая строго больше области измерений. Эта неопределённость обычно приводит к различного рода проблемам, некоторые из которых мы рассмотрим в следующем разделе. В-третьих, карты коррекции поверхности обычно генерируются из зашумленных данных, что делает необходимой предварительную фильтрацию, (i) чтобы избежать ухудшения коррекции из-за влияния ошибок измерений и (ii) чтобы упростить расчёт карт травления. В идеале схема фильтрации данных измерений должна учитывать по крайней мере два аспекта: (i) характеристики измерительного инструмента и (ii) низкую эффективность травления для мелкомасштабных (высокочастотных) особенностей. В настоящей статье мы не рассматриваем первый аспект, поскольку он, хотя и представляет большой интерес сам по себе, тематически довольно далек от вопросов, анализируемых здесь. Напротив, второй аспект оказывается чрезвычайно важным, поскольку травление мелкомасштабных особенностей, которые могут возникать из-за шума, не только нежелательно, но и чрезвычайно неэффективно.

В данной работе предлагается математический аппарат, позволяющий изучать проблему расчёта карты травления в широком контексте, применимый к таким областям, как обратные задачи в электростатике или реконструкция изображений. Будет показано, что особенно непрерывные модели хорошо подходят для их изучения средствами функционального анализа. Мы рассматриваем математические модели различной степени сложности, дискретные и непрерывные, от простых одномерных до более реалистичных двумерных. В разделе \ref{s:FuncAna} обсуждается формулировка проблемы расчёта карты травления с точки зрения функционального анализа, что приводит нас к рассмотрению бесконечномерных линейных уравнений. В разделе \ref{s:EigsAAstar} мы анализируем собственные значения и собственные функции оператора, связанного с упомянутыми выше линейными уравнениями. Поскольку в практических приложениях преимущественно используется поточечное описание данных, в разделе \ref{s:RKHS} мы показываем, как решить задачу расчёта времени экспозиции в рамках гильбертовых пространств с воспроизводящими ядрами. В разделе \ref{s:Filtering} мы обсуждаем некоторые следствия нашей работы для фильтрации данных, используемых в качестве входных для расчёта времени экспозиции. В разделе \ref{s:conclusion} мы рассматриваем основные результаты нашей работы и делаем некоторые заключительные замечания. Для удобства читателя мы добавили приложение с описанием методов функционального анализа, используемых в настоящей статье.

\section{Формулировка задачи в рамках функционального анализа и теории операторов}
\label{s:FuncAna}

\subsection{Предварительные замечания}
\label{eq:Preliminary}

Мы начинаем с простого одномерного примера, который позволяет нам объяснить суть подхода, который мы собираемся использовать, и рассмотреть обе упомянутые ранее проблемы: корректности задачи и вычисления времени экспозиции вне области измерений.

Предполагается, что профиль коррекции поверхности $h(x)$ задан на интервале $[-L,L]$, где $L > 0$. Мы также предполагаем, что функция пучка задана гауссовой функцией с шириной $\sigma$
\begin{equation}
\label{eq:Gaussian1}
f (x; \sigma) = \frac{1}{\sqrt{2\pi \sigma ^2 } } e^{-\frac{x^2 }{2 \sigma ^2} } .
\end{equation}
Требуется определить карту травления $t(x)$ таким образом, чтобы свёртка $f (\cdot; \sigma) \ast t(\cdot)$ была как можно ближе к $h(x)$ на интервале $[-L,L]$.
Один из способов достижения этой цели — сформулировать уравнение следующего вида
\begin{equation}
\label{eq:convolution1}
h(x) = \int_{-\infty }^{\infty } f (x - x^{\prime } ) t (x^{\prime } ) \, dx^{\prime } , \qquad x \in [-L,L] ,
\end{equation}
где, в отличие от уравнений свёртки стандартного типа, см., например, \cite{HirschWidd}, равенство выполняется только на подмножестве вещественной оси, интервале $[-L, L]$.
В то же время, время экспозиции $t(x)$ должно быть задано на области, большей, чем $[-L,L]$, из-за ненулевого размера функции пучка $f (x)$. Для простоты мы предполагаем, что
функция $t(x)$ определена на всей вещественной оси $\mathbb{R}$.

Эта несовместимость областей определения между требуемым профилем коррекции поверхности $h(x)$ и временем экспозиции $t(x)$
не просто неудобство, а значительная проблема, причем в нескольких отношениях. Во-первых, её можно рассматривать как проблему согласованного задания значений.
Действительно, если мы стремимся распространить уравнение (\ref{eq:convolution1}) на всю вещественную ось, мы должны задать значения $h(x)$ вне интервала $[-L,L]$.
Априори на этой области мы можем приписать $h(x)$ любые значения, определяя таким образом расширение $h_{ext} (x)$ функции $h(x)$. Расширенный профиль коррекции поверхности $h_{ext} (x)$
может затем быть использован для вычисления карты травления $t^{\ast } (x)$ с помощью некоторых стандартных методов, после чего может быть вычислен профиль коррекции поверхности $h^{\ast } (x)$
как свёртка $t^{\ast } (x)$ с функцией пучка $f (x)$. Для функций пучка
типа (\ref{eq:Gaussian1}), в силу нелокальной природы операции свёртки, значения $h^{\ast } (x)$ внутри интервала $[-L,L]$ (точнее, в некоторой окрестности
его границ, размер которой пропорционален $\sigma $), будут зависеть от значений, приписанных $h_{ext} (x)$ снаружи. Для согласованности, внутри интервала $[-L,L]$ заданный
профиль коррекции поверхности $h_{ext} (x)$ должен совпадать с вычисленным $h(x)$.
Следовательно, мы не свободны в определении расширения $h_{ext} (x)$, и, более того, не существует очевидного явного правила или процедуры для задания такого расширения.
Во-вторых, несовместимость областей приводит к практической проблеме, которую можно резюмировать следующим образом. Чтобы задать процедуру расширения и обеспечить 
конечность времени обработки,
иногда предполагают, что расширенная функция $h_{ext} (x)$ обращается в ноль вне некоторого расширенного интервала, который строго больше, чем $[-L,L]$.
Тогда вблизи границ интервала $[-L,L]$ обычно наблюдаются расхождения между заданным $h(x)$ и вычисленным $h^{\ast } (x)$, которые становятся всё более
заметными по мере роста градиента $h(x)$.

Дальнейшая трудность, а именно некорректность постановки задачи, возникает при решении (\ref{eq:convolution1}) и сохраняется, даже если пренебречь указанной выше проблемой 
несовместимости областей определения.
Предположим, что профиль коррекции поверхности задан на всей вещественной оси. Таким образом, нам нужно будет решить довольно стандартное уравнение свёртки
$h(x) = \left( f (\cdot; \sigma) \ast t(\cdot) \right) (x) $, $x\in \mathbb{R}$.
Сразу отметим, что если положить $\sigma ^2 = 2\tau $ в (\ref{eq:Gaussian1}), функция влияния инструмента становится равной ядру теплопроводности на вещественной оси
\begin{displaymath}
\phi _{\tau } (x) = \frac{1}{\sqrt{4\pi \tau } } e^{-\frac{x^2 }{4 \tau } } ,
\end{displaymath}
так что свёртка $u_{\tau} = \phi _{\tau } \ast t $ фактически является решением уравнения теплопроводности
\begin{displaymath}
\partial _{\tau } u_{\tau} = \partial _x^2 u_{\tau} ,
\end{displaymath}
с начальным условием $u_{0} (x) = t(x)$. Таким образом, решение уравнения свёртки эквивалентно рассмотрению обратных решений уравнения
теплопроводности\footnote{Заметим, что выполняя виковский поворот и делая временную переменную $\tau $ мнимой как $2\tau = i \eta $, мы получаем
ядро, описывающее распространение электромагнитного излучения в параксиальном приближении.}. Это
яркий пример некорректно поставленной задачи в смысле Адамара, и к ней нельзя подступиться стандартными методами. Действительно, решения уравнения теплопроводности
хорошо известны своими сглаживающими свойствами. Это означает, в свою очередь, что с ростом $\tau $ обратные решения быстро теряют гладкость.

Также стоит упомянуть, что свёртка $u_{\tau} = \phi _{\tau } \ast t $ тесно связана с преобразованием Вейерштрасса, определяемым как $\phi _1 \ast t $, см., например,
\cite{BrychkovPrudnikov}. В принципе, чтобы получить $t(x)$ из $u_{\tau }$, можно было бы поступить так же, как при обращении преобразования Вейерштрасса.
Однако оказывается, что ввиду упомянутой потери гладкости, для корректного математического рассмотрения этой задачи требуется использовать аппарат
обобщённых функций: это означает, что решения уравнения свёртки даже для гладких $u_{\tau}$ являются обобщёнными функциями, которые не могут быть легко представлены
численно, не говоря уже об их использовании в качестве карт травления.

\subsection{Задача расчета карты травления как бесконечномерное линейное уравнение}
\label{ss:InfiniteDimLinEq}

Как мы видели выше, использование уравнения свёртки (\ref{eq:convolution1}) без указания строгих математических условий, в которых оно должно формулироваться,
приводит к различным неоднозначностям и проблемам. Один из возможных подходов к формализации задачи расчёта времени экспозиции состоит в использовании инструментов,
предлагаемых функциональным анализом. Он реализуется следующим образом: вместо рассмотрения уравнения (\ref{eq:convolution1}) мы рассматриваем как $h(x)$, так и $\left( f \ast t \right) (x) $,
ограниченные на $[-L,L]$, как элементы некоторого нормированного векторного пространства функций на $[-L,L]$ и требуем их равенства. При этом мы предпочитаем задавать математическую
структуру как можно более обще, в то же время стараясь избегать ненужных усложнений. Здесь нам кажется удобным рассматривать
функции $h({\bm x})$ от ${\bm x} \in \Omega \subset \mathbb{R}^n $ как элементы гильбертова пространства $L^2 \left( \Omega , \lambda \right)$ функций на компактном множестве $\Omega $,
которые мы также будем называть функциями измерений или функциями коррекции, фиксированную функцию $f({\bm x})$, которая является функцией инструмента (пучка),
и функции $t({\bm x})$ из гильбертова пространства $L^2 \left( \mathbb{R}^n , \lambda \right)$ функций на $\mathbb{R}^n $, которые мы называем восстанавливающими функциями.
Мера $\lambda $ для простоты предполагается инвариантной относительно сдвигов мерой Лебега на $\mathbb{R}^n$. Свёртка определяет оператор
\begin{displaymath}
f \ast t = A:L^2 \left( \mathbb{R}^n , \lambda \right) \longrightarrow L^2 \left( \Omega , \lambda \right) ,
\end{displaymath}
который мы предполагаем ограниченным для всех интересующих случаев. Достаточное, но не необходимое условие для этого даётся неравенством Юнга, если мы предположим, что
$\Vert f \Vert _{L^1 \left( \mathbb{R}^n , \lambda \right)} < \infty $. Более того, в этом случае мы можем положить для
простоты $\Vert f \Vert _{L^1 \left( \mathbb{R}^n , \lambda \right)} = 1 $.
Тогда проблема определения восстанавливающей функции $t$ принимает форму бесконечномерного линейного уравнения
\begin{equation}
\label{eq:Ax=b}
At^{\star } = h  ,
\end{equation}
где $h$ — заданный результат измерения. Как и в конечномерном случае, это уравнение может быть решено с помощью псевдообратного оператора Мура–Пенроуза
$A^{\dagger } $, при условии
$h \in D(A^{\dagger }) = R(A) \oplus N(A^{\ast })$, см. \cite{BenisraelGreville,CheverdaKostin} (Здесь $R(A)$ — образ оператора $A$, а $N(A^{\ast })$ — ядро
сопряжённого оператора $A^{\ast }$. Более подробное изложение можно найти в разделе \ref{app:FuncAna}.). Эти решения совпадают с экстремальными решениями, полученными минимизацией
величины $ \Vert h - At \Vert _{L^2 \left( \Omega , \lambda \right)} $, где норма $ \Vert t \Vert _{L^2 \left( \mathbb{R}^n , \lambda \right)} $ функции $t$ достигает своего минимума.

Если вместо решения уравнения (\ref{eq:Ax=b}) мы минимизируем целевую функцию
\begin{eqnarray}
\label{eq:QuadMeritFunction}
\Phi (\{ t \} ) &=& \frac{1}{2} \, \Vert h - At \Vert _{L^2 \left( \Omega , \lambda \right)}^2  = \nonumber
\\
& & \frac{1}{2} \langle A^{\ast } At ,  t \rangle _{L^2 \left( \mathbb{R}^n , \lambda \right)} - \langle A^{\ast } h, t \rangle  _{L^2 \left( \mathbb{R}^n , \lambda \right)} +
\frac{1}{2} \, \Vert h \Vert _{L^2 \left( \Omega , \lambda \right)}^2 ,
\end{eqnarray}
то получим вариационное (уравнение Эйлера)
\begin{equation}
\label{eq:Euler}
A^{\ast } \! A\, t^{\star \star } = A^{\ast }  h .
\end{equation}
Здесь $A^{\ast }$ обозначает сопряжённый оператор к $A$
\begin{displaymath}
A^{\ast }  h ({\bm x}) = \int_{\Omega } f({\bm x} - {\bm x}^{\prime } ) h ({\bm x}^{\prime } ) \, d^n {\bm x}^{\prime } .
\end{displaymath}
Уравнение (\ref{eq:Euler}) может быть решено путём обращения оператора $A^{\ast } A$ на его
образе\footnote{Оператор $A$ не является нормальным (более подробное описание дается в разделе  \ref{app:FuncAna}) из-за 
проблемы несовместимости областей
упомянутой в разделе \ref{eq:Preliminary}. То есть $A^{\ast } A \neq A A^{\ast } $, поскольку $A$ действует из
$L^2 \left( \mathbb{R}^n , \lambda \right)$ в $L^2 \left( \Omega , \lambda \right)$.}. В то же время,
оператор $A^{\ast } A$ может быть использован для построения псевдообратного оператора $A^{\dagger } $ для $A$. С этой целью мы предполагаем, что $A$ является
компактным оператором. Как мы увидим позже, это предположение выполняется для всех приложений, которые мы будем рассматривать.

Тогда интегральный оператор
\begin{displaymath}
\left( A^{\ast }  A \right) t ({\bm x}^{\prime }) = \int_{\mathbb{R}^n } k_{A^{\ast }  A} ( {\bm x}^{\prime }, {\bm x}^{\prime \prime } ) t({\bm x}^{\prime \prime } ) \, d^n {\bm x}^{\prime \prime } ,
\end{displaymath}
является положительным и компактным, с ядром, заданным формулой
\begin{equation}
\label{eq:IntegralForKernel}
k_{A^{\ast }  A} ( {\bm x}^{\prime } , {\bm x}^{\prime \prime } ) =  \int_{\Omega } f ({\bm x}^{\prime } - {\bm x} )
f ({\bm x} - {\bm x}^{\prime \prime }) \, d^n {\bm x} .
\end{equation}
Его спектр $\Sigma $, следовательно, дискретен
$\Sigma = \{ \lambda _n \} _{n \in \mathbb{N} }$, положителен $\lambda _n \geq 0$ и ограничен сверху. Пусть
$t_n \in L^2 \left( \mathbb{R}^n , \lambda \right)$ — соответствующие собственные функции. Тогда $t_n $ отождествляются с правыми
сингулярными векторами оператора $A$, тогда как левые сингулярные векторы $h_n$ задаются формулой
\begin{equation}
\label{eq:LeftSingVec}
L^2 \left( \Omega , \lambda \right)  \ni h_n  = \frac{1}{\sqrt{\lambda _n }} At_n .
\end{equation}
Сингулярные значения оператора $A$ равны $\sqrt{\lambda _n }$, так что
\begin{equation}
\label{eq:singVectors}
A t_n = \sqrt{\lambda _n } h_n  , \qquad  A^{\ast }  h_n = \sqrt{\lambda _n } t_n .
\end{equation}
Для $h \in R(A) \oplus N(A^{\ast }) \subset L^2 \left( \Omega , \lambda \right)$ псевдообратный оператор к $A$ может быть вычислен как
\begin{equation}
\label{eq:pseudoinverse}
A^{\dagger } h = \sum_{\lambda _n \neq 0} \frac{1}{\lambda _n } \langle t_n , A^{\ast} h \rangle  _{L^2 \left( \mathbb{R}^n , \lambda \right)} \, t_n =
\sum_{\lambda _n \neq 0} \frac{1}{\sqrt{\lambda _n } } \langle h_n , h \rangle  _{L^2 \left( \Omega , \lambda \right)} \, t_n  .
\end{equation}
Проблема решения бесконечномерного уравнения (\ref{eq:Ax=b}) может, таким образом, быть решена путём вычисления спектра и
собственных функций оператора $ A^{\ast } A$ с ядром (\ref{eq:IntegralForKernel}).
С другой стороны, из уравнений (\ref{eq:singVectors}) следует, что $\lambda _n$ и $h_n$ являются собственными значениями и собственными функциями
оператора $ A A^{\ast }$, действующего на $ L^2 \left( \Omega , \lambda \right)$ как
\begin{equation}
\label{eq:dualOperator}
\left( A A^{\ast }  \right) h ({\bm x}^{\prime }) =  \int_{\Omega }  k_{AA^{\ast } } ( {\bm x}^{\prime }, {\bm x}^{\prime \prime } ) h({\bm x}^{\prime \prime } ) \, d^n {\bm x}^{\prime \prime }  =
\int_{\Omega }  (f \ast f) ({\bm x}^{\prime } - {\bm x}^{\prime \prime }) h({\bm x}^{\prime \prime } ) \, d^n {\bm x}^{\prime \prime } ,
\end{equation}
с трансляционно инвариантным ядром $ k_{AA^{\ast } } ( {\bm x}^{\prime }, {\bm x}^{\prime \prime } ) = (f \ast f) ({\bm x}^{\prime } - {\bm x}^{\prime \prime })$.

Из уравнений (\ref{eq:pseudoinverse}) следует, что сингулярные значения $\sqrt{\lambda _n }$, грубо говоря, представляют собой эффективность, с которой восстанавливающая функция $t_n$
реализуется в результате коррекции $h_n$, поскольку в выражении после второго знака равенства, время экспозиции, соответствующее
собственной функции $t_n$, должно быть усилено с множителем
$\frac{1}{\sqrt{\lambda _n }}$. Для коррекции поверхности посредством ионного травления это означает, что времена экспозиции, основанные на собственных векторах с более высокой
эффективностью травления (большими сингулярными значениями), оказывают большее влияние на изменение профиля поверхности, чем те, которые соответствуют меньшим сингулярным значениям.

Резюмируя, можно сказать, что с концептуальной точки зрения введение псевдообратного оператора Мура–Пенроуза позволяет нам понять
задачу решения бесконечномерного уравнения (\ref{eq:Ax=b}) по аналогии с хорошо известной задачей решения систем линейных уравнений типа $Ax=b$.
Однако заметим, что в случаях, когда $ A^{\ast } A$ не имеет спектральной щели — в противном случае он имеет конечный ранг и рассматриваемая проблема сводится
к конечномерной — возникает несколько аспектов, которые необходимо принять во внимание.
Во-первых, в практических приложениях бесконечную сумму в (\ref{eq:pseudoinverse}) приходится усекать, если только нам не удаётся найти аналитическое выражение
для псевдообратного оператора. Какой порядок усечения выбрать, не очевидно и сильно зависит от рассматриваемой задачи.
Во-вторых, решения уравнения (\ref{eq:Ax=b}) оказываются по своей природе неустойчивыми, так что необходимо введение некоторого рода регуляризации.
Обычно вместо непосредственного решения уравнения (\ref{eq:Ax=b}) удобнее решать усечённый и регуляризованный вариант
задачи оптимизации (\ref{eq:QuadMeritFunction}).

\subsection{Примеры} 

Здесь мы представим несколько примеров прикладных задач, которые могут быть рассмотрены с помощью формализма, представленного выше.

{\bf Пример 1: Гауссово распределение.} Основной пример функции инструмента — это $n$-мерное гауссово распределение
\begin{displaymath}
f ({\bm x} ; B) = \frac{1}{\sqrt{(2\pi )^n \vert \mathrm{det} B\vert }} e^{-\frac{1}{2} \langle B^{-1} {\bm x} , {\bm x} \rangle _{\mathbb{R}^n }} .
\end{displaymath}
Для $n=2$ это распределение используется для моделирования функций пучка при вычислении карты травления для заданной
коррекции профиля поверхности, заданной на области $\Omega$.
На практике ионный пучок настраивается таким образом, чтобы $f ({\bm x} ; B)$ являлась осесимметричной функцией
\begin{displaymath}
B = \left( \begin{array}{c c} \sigma ^2 & 0 \\ 0 & \sigma ^2 \end{array} \right) .
\end{displaymath}
В данной работе мы будем рассматривать в качестве примеров случаи одно- и двумерного гауссова распределения
$(2\pi )^{-\frac{n}{2}} e^{-\frac{\vert {\bm x} ^2 }{2 \sigma ^2 }} $, $n = 1, 2$. В случаях, когда распределение плотности ионов в ионном пучке
отклоняется от гауссова, для улучшения математического моделирования пучка может быть использовано разложение Эджворта \cite{BlinMoes}.

{\bf Пример 2: Распределение Коши.} Менее очевидный пример — распределение
\begin{displaymath}
f ({\bm x} ; \sigma) = \frac{c_n \sigma }{(\sigma ^2 + \vert {\bm x} \vert ^2)^{\frac{n+1}{2}} }  , \qquad c_n = \frac{\Gamma \left( \frac{n + 1}{2} \right) }{\pi ^{\frac{n + 1}{2}}} .
\end{displaymath}
В одном измерении $f ({\bm x} ; \sigma)$ совпадает с распределением Коши. Для произвольного $n \in \mathbb{N} $ функция $f ({\bm x} ; \sigma)$ является
ядром Пуассона для уравнения Лапласа в верхнем полупространстве
$H^{+} = \{ ({\bm x}, y) : {\bm x} \in \mathbb{R}^n , y>0 \} \subset \mathbb{R}^{n+1}$. Таким образом, уравнение (\ref{eq:Ax=b}) в этом случае описывает проблему
восстановления граничного условия для уравнения Лапласа
 на $H^{+}$ по заданному распределению потенциала $h({\bm x}) $ на области $\Omega \subset H^{+}$ на расстоянии $y = \sigma $ от плоскости $y = 0$.

Оба случая, распределения Гаусса и распределения Коши (ядро Пуассона в случае нескольких переменных) могут быть обобщены с помощью ядра дробного уравнения теплопроводности
\begin{displaymath}
\partial _{\tau } u_{\tau } + (-\Delta )^s u = 0 ,
\end{displaymath}
на $\mathbb{R}^n $. Гауссов случай задаётся $s = 1$, тогда как для распределения Коши $s = \frac{1}{2}$.  Для общих $0 < s < 1$, могут быть даны верхние и нижние оценки значений ядра (см. \cite{Vazquez}).

{\bf Пример 3: Скользящее среднее.} Пример, который ближе к обработке изображений, возникает, когда оператор $A$ описывает усреднение по окрестности оцениваемой
координаты ${\bm x}$. Обычно усреднение происходит либо по шару радиуса $\sigma $
\begin{displaymath}
f ({\bm x} ; \sigma ) = \frac{1}{\mathrm{Vol} (D_{\sigma }) } \chi _{D_{\sigma }} ({\bm x}) , \qquad D_{\sigma } = \{ {\bm x} \in \mathbb{R}^n : \vert {\bm x} \vert \leq \sigma \} \subset \mathbb{R}^n ,
\end{displaymath}
либо по кубу со стороной $2\sigma $. В этом случае уравнение (\ref{eq:Ax=b}) соответствует проблеме восстановления исходного изображения по его усреднённой версии, известной на области $\Omega $.

{\bf Пример 4: Спектральное усечение.} До сих пор у нас были только положительные функции влияния, то есть $f ({\bm x}) \geq 0$ для всех ${\bm x} \in \mathbb{R}^n $. Функция
\begin{displaymath}
f({\bm x} ; K) = \left( \frac{4K}{2 \pi } \right)^n \prod_{i=1}^n \mathrm{sinc} \, K x_i  ,
\end{displaymath}
которая в пространстве Фурье удаляет вклады волновых чисел $\vert k_i \vert > K$ не является знакоопределённой.  Заметим, что соответствующий оператор $A$ ограничен в силу теоремы Парсеваля (тогда как его $L^1$-норма не конечна). Мы
выбрали множитель таким образом, чтобы сделать операторную норму $A$ равной единице.

{\bf Пример 5: Восстановление электростатического поля на границе бесконечной трубки.} Теперь приведём более сложный пример восстановления электростатического потенциала. Рассмотрим бесконечную трубку радиуса $R$ и предположим, что осесимметричный потенциал на границе этой трубки равен $\phi(R, x)$, $x \in \mathbb{R}$. Тогда потенциал внутри трубки на радиусе $r$ может быть вычислен как
\begin{displaymath}
\phi (r, x) = \int_{\mathbb{R}} K_R (r, x - x^{\prime } ) \phi (R, x^{\prime } ) \, dx^{\prime } , \qquad r < R ,
\end{displaymath}
где ядро Пуассона $K_R (r, \xi )$ задаётся формулой
\begin{displaymath}
K_R (r, \xi ) = \frac{1}{2\pi } \int_{\mathbb{R}} \frac{I_0 (kr)}{I_0 (kR)} e^{ik \xi } \, dk ,
\end{displaymath}
см. \cite{Bertram}.
Тогда, предполагая, что мы знаем потенциал, скажем, при $r = 0$ на интервале $[-L,L]$, так что $h(x) =  \phi (r, x)$ и полагая $t(x^{\prime } ) = \phi (R, x^{\prime } )$
уравнение (\ref{eq:Ax=b}) соответствует проблеме
восстановления граничного потенциала на границе трубки по потенциалу, частично известному вдоль оси. Такие вопросы могут возникать, например, когда мы пытаемся
оптимизировать форму электростатических линз, чтобы минимизировать аберрации в электронно-оптических системах.

Отметим, что функции влияния в приведённых выше примерах демонстрируют различные типы убывания на бесконечности, от полиномиального (в примерах 2 и 4), экспоненциального (в примере 5) до убывания быстрее экспоненциального (в примерах 1 и 3).

\subsection{Свойства операторов $A$ и $A^{\ast } A $}
\label{ss:PropertiesAAStar}

В этом разделе мы устанавливаем основные математические свойства операторов $A$ и $A^{\ast } A $, необходимые для построения псевдообратного оператора для $A$ в
разделе \ref{ss:InfiniteDimLinEq}, используя оценки с помощью различного вида норм. В частности, мы покажем, что в определённом смысле $A^{\ast } A $ можно рассматривать как бесконечномерную матрицу. Заметим, что определения и свойства
норм, используемых в этом разделе, можно найти в разделе \ref{app:FuncAna}.

Как мы уже отмечали ранее, для всех функций инструмента из примеров 1--5 оператор $A$ ограничен. Более точно, он компактен, что можно показать, используя
критерии компактности из главы IX в \cite{KantAkil} (более подробная информации дана в разделе \ref{app:FuncAna}. Это обосновывает  предположение о компактности, сделанное в разделе \ref{ss:InfiniteDimLinEq}.

Мы начнём с того, что квадрат нормы Гильберта–Шмидта оператора $A^{\ast } A$ задаётся выражением
\begin{displaymath}
\Vert A^{\ast } A \Vert_{HS}^2 = \int_{\mathbb{R}^n } \int_{\mathbb{R}^n } \left( k ( {\bm x}^{\prime } , {\bm x}^{\prime \prime } ) \right)^2 \, d^n {\bm x}^{\prime } d^n {\bm x}^{\prime \prime } =
\int_{\Omega } \int_{\Omega }\left( (f \ast f) ({\bm x} - {\bm x}^{\prime } ) \right)^2 \, d^n {\bm x} d^n {\bm x}^{\prime }   ,
\end{displaymath}
которое конечно, если $(f \ast f) ({\bm x})$ либо непрерывна, либо ограничена на $\mathbb{R}^n$, в этом случае
\begin{equation}
\label{ineq:HS}
\Vert A^{\ast } A \Vert_{HS}^2 \leq  \left( \mathrm{Vol} (\Omega ) \right)^2 \Vert (f \ast f)^2 \Vert _{L^{\infty } (\Omega )} < \infty  ,
\end{equation}
так что $A^{\ast } A$ является самосопряжённым (поскольку $A^{\ast } A$ симметричен и ограничен) оператором Гильберта–Шмидта. Заметим, что оценки нормы Гильберта–Шмидта типа (\ref{ineq:HS})
могут быть использованы для получения границ на норму Шаттена--фон Неймана оператора $ A^{\ast } A$
\begin{displaymath}
\sum_{n=1}^{\infty } \lambda _n^2  = \Vert A^{\ast } A \Vert_{HS}^2  = \int_{\Omega } \int_{\Omega }\left( (f \ast f) ({\bm x} - {\bm x}^{\prime } ) \right)^2 \, d^n {\bm x} d^n {\bm x}^{\prime } .
\end{displaymath}
Ядерная норма оператора $A^{\ast } A$ задаётся выражением
\begin{displaymath}
\int_{\mathbb{R}^n } k ( {\bm x} , {\bm x} )  \, d^n {\bm x} =  \int_{\mathbb{R}^n } \int_{\Omega } f({\bm x} - {\bm x}^{\prime } ) f({\bm x} - {\bm x}^{\prime } ) \, d^n {\bm x} \, d^n {\bm x}^{\prime }  =
\mathrm{Vol} (\Omega ) \Vert f \Vert ^2_{L^2 \left( \mathbb{R}^n , \lambda \right)} .
\end{displaymath}
Следовательно, $A^{\ast } A$ является ядерным оператором, если $\Vert f \Vert _{L^2 \left( \mathbb{R}^n , \lambda \right)} < \infty $ и
\begin{equation}
\label{eq:traceNorm}
\sum_{n=1}^{\infty } \lambda _n  = \mathrm{Vol} (\Omega ) \Vert f \Vert ^2_{L^2 \left( \mathbb{R}^n , \lambda \right)}  .
\end{equation}
Вышеприведённое обсуждение различных норм оператора $A^{\ast } A$ может быть резюмировано следующим образом. Во-первых, из соотношений (\ref{ineq:HS}) и (\ref{eq:traceNorm})
мы можем получить некоторую информацию (в смысле $L^p$-норм) о собственных значениях $A^{\ast } A$. Во-вторых, как объясняется, например, в \cite{BrianDavies}, ядерные операторы, одним из которых
оказывается $A^{\ast } A$, являются следующим в иерархии классом подлинно бесконечномерных операторов, включающим в себя операторы конечного ранга, которые в контексте функционального анализа являются просто конечномерными матрицами.
В этом смысле $A^{\ast } A$ близок к матрице и в некоторых случаях может быть эвристически понят как таковая.

Наибольшее собственное значение $A^{\ast } A$ может быть оценено с помощью вариационной теоремы для $\Vert t \Vert_{L^2 (\mathbb{R} )} = 1$ как
\begin{eqnarray*}
\langle A^{\ast } A t, t \rangle _{L^2 \left( \mathbb{R}^n , \lambda \right)} & = & \Vert f \ast t \Vert_{L^2 \left( \Omega , \lambda \right)}^2 \leq \Vert f \ast t \Vert_{L^2 \left( \mathbb{R}^n , \lambda \right)}^2 \leq
\\
& &
\Vert f \Vert_{L^1 \left( \mathbb{R}^n , \lambda \right)}^2 \Vert t \Vert_{L^2 \left( \mathbb{R}^n , \lambda \right)}^2 = \Vert f \Vert_{L^1 \left( \mathbb{R}^n , \lambda \right)}^2 = 1 ,
\end{eqnarray*}
где мы использовали неравенство Юнга для оценки свёртки, предполагая, что $L^1 \left( \mathbb{R}^n , \lambda \right)$-норма
функции инструмента равна единице.
Таким образом, наибольшее собственное значение $\lambda _1$ удовлетворяет неравенству $0 < \lambda _1 \leq 1$. Более точно, неравенство $\lambda _1 < 1$ строгое, поскольку
само неравенство Юнга насыщается для гауссовых $f$ и $t$ (см., например, \cite{BrascampLieb}), для которых первое неравенство в вышеприведённой формуле опять таки 
было бы строгим ($\Vert f \ast t \Vert_{L^2 ([-L,L])}^2 < \Vert f \ast t \Vert_{L^2 (\mathbb{R} )}^2$). Такая же
оценка может быть получена для функции инструмента спектрального усечения из примера 4. Заметим, что для 
положительных функций инструмента 
$f({\bm x}) \geq 0$ первая собственная функция $t_1$ может быть охарактеризована с помощью теоремы Крейна–Рутмана.
А именно, в этом случае выполняется неравенство $t_1 \geq 0$.

Утверждения о свойствах собственных значений $A^{\ast } A$, которые мы привели до этого момента, были получены общими 
методами, которые справедливы для широких
классов функций влияния инструмента. Поэтому неудивительно, что количество информации, содержащейся в них, не позволяет провести детальную характеристику
собственных значений. Эта ситуация типична для подходов, основанных на оценках норм. Более детальное исследование спектра $A^{\ast } A$ будет представлено в разделе \ref{s:EigsAAstar}.

Теперь оценим свойства ядра оператора $A^{\ast } A$. Для $t \in  N(A^{\ast } A) $ мы имеем соотношение
\begin{displaymath}
\langle A^{\ast } A t, t \rangle _{L^2 (\mathbb{R}^n )} = \Vert f \ast t \Vert_{L^2 \left( \Omega , \lambda \right)}^2 = 0 .
\end{displaymath}
Это эквивалентно $f \ast t = 0 $ почти всюду на $\Omega$. Заметим, что для некоторых классов функций влияния (например, гауссовой $f$, или $f$, чьё преобразование Фурье имеет конечный носитель, как в примере 4) функция $f \ast t$ будет аналитической, так
что из её равенства нулю на $\Omega$ следует, что $f \ast t = 0 $ на $\mathbb{R}^n$. Если, как в случае гауссовой $f$, преобразование
Фурье $\mathcal{F} f$ не равно нулю почти всюду, то $t =0$ в смысле $ L^2 (\mathbb{R}^n )$.
Это означает, что ядро $A^{\ast } A$ в этом случае тривиально, то есть $A^{\ast } A$ не имеет спектрального зазора, другими словами $\inf _n \lambda _n = 0$. Из отсутствия спектральной щели, в свою очередь, следует, что уравнение (\ref{eq:Euler}) некорректно поставлено и нуждается в
регуляризации, см. \cite{BenningBurger}. В противоположность этому, для функций инструмента в примере 4, $A^{\ast } A$ имеет нетривиальное ядро, состоящее из функций, носитель которых в пространстве Фурье лежит за пределами усекающего волнового числа $K$.

Наконец, заметим, что хотя резольвенты операторов $A^{\ast } A$ и $A A^{\ast }$ могут быть представлены с помощью ряда Лиувилля–Неймана, мы не нашли явных выражений, которые могли бы быть использованы для проведения детального анализа их спектров.
Мы лишь отметим, что их соответствующие итерированные ядра задаются формулами
\begin{eqnarray*}
k_{A^{\ast } A}^{(n)} ( {\bm x}^{\prime }, {\bm x}^{\prime \prime } ) = & & \\
\int_{\Omega } ... \int_{\Omega } f ({\bm x}^{\prime } - {\bm x}_1) & & 
(f \ast f) ({\bm x}_1 - {\bm x}_2) ... (f \ast f) ({\bm x}_{n-1} - {\bm x}_{n}) f ({\bm x}_{n} - {\bm x}^{\prime \prime })  \, d^n {\bm x}_1 ... d^n {\bm x}_{n} ,
\end{eqnarray*}
и
\begin{displaymath}
k_{AA^{\ast } }^{(n)} ( {\bm x}^{\prime }, {\bm x}^{\prime \prime } ) =
\int_{\Omega } ... \int_{\Omega } (f \ast f) ({\bm x}^{\prime } - {\bm x}_1) ... (f \ast f) ({\bm x}_{n-1} - {\bm x}^{\prime \prime }) \, d^n {\bm x}_1 ... d^n {\bm x}_{n-1} .
\end{displaymath}

\subsection{Постановка задачи в диcкретном случае}

В дискретном случае мы отождествляем карты травления с гильбертовым пространством $l^2 (\Delta \mathbb{Z}^n )$ абсолютно квадратично-суммируемых $n$-мерных
последовательностей $ (\alpha )_{i_{1}...i_{n}} $ как
\begin{equation}
\label{eq:discreteT}
t ({\bm x}) = \sum_{(i_{1},...,i_{n}) \in \mathbb{Z}^n}  \alpha _{i_{1}...i_{n}} \delta _{ \Delta (i_{1},...,i_{n}) } ({\bm x}) .
\end{equation}
Оператор $A$ определяется как
\begin{equation}
\label{eq:discreteH}
\left( A (\alpha )_{i_{1}...i_{n}} \right) ({\bm x} ) = \sum_{(i_{1},...,i_{n}) \in \mathbb{Z}^n}  \alpha _{i_{1}...i_{n}} f ({\bm x} - \Delta (i_{1},...,i_{n}) ) ,
\end{equation}
а оператор $A^{\ast } A$ может быть понят как бесконечномерная матрица
\begin{equation}
\label{eq:discreteKernelAAstar}
(A^{\ast } A)_{(i_{1},...,i_{n}), (j_{1},...,j_{n})} = \int_{\Omega } f (\Delta (i_{1},...,i_{n}) - {\bm x} ) f ({\bm x} - \Delta (j_{1},...,j_{n}) ) \, d^n {\bm x} .
\end{equation}
Оператор $AA^{\ast }$, в свою очередь, определяется своим интегральным ядром
\begin{displaymath}
k_{AA^{\ast }} ({\bm x} , {\bm x}^{\prime } ) = \sum_{(i_{1},...,i_{n}) \in \mathbb{Z}^n}  f ({\bm x} - \Delta (i_{1},...,i_{n}) ) f (\Delta (i_{1},...,i_{n}) - {\bm x}^{\prime } ) .
\end{displaymath}
Для гауссовой функции инструмента это ядро может быть явно определено через выражение
\begin{displaymath}
\sum_{n \in \mathbb{Z} } \frac{1}{\sqrt{2 \pi \sigma ^2 }} e^{ -\frac{(x - \Delta n)^2}{2\sigma ^2 } } \frac{1}{\sqrt{2 \pi \sigma ^2 }}
e^{ -\frac{(\Delta n - x^{\prime } )^2}{2\sigma ^2 } } = \frac{1}{2 \pi \sigma ^2 }
e^{ -\frac{x^2 + x^{\prime \, 2} }{2\sigma ^2 } } \, \theta _3 \left( \frac{\Delta (x + x^{\prime } ) }{2 i \sigma } , \frac{i \Delta ^2 }{\pi \sigma ^2 } \right) ,
\end{displaymath}
где $\theta _3 $ — одна из тета-функций Якоби, см. \cite{WhittakerWatson}. Тогда ядро $AA^{\ast }$ может быть записано как
\begin{equation}
\label{eq:discreteKernelAstarA}
k_{AA^{\ast }} ({\bm x} , {\bm x}^{\prime } ) = \left( \frac{1}{2 \\pi \sigma ^2 } \right)^n \prod_{m=1}^n
e^{ -\frac{x_m^2 + x_m^{\prime \, 2} }{2\sigma ^2 } } \, \theta _3 \left( \frac{\Delta (x_m + x_m^{\prime } ) }{2 i \sigma } , \frac{i \Delta ^2 }{\pi \sigma ^2 } \right) .
\end{equation}
Для произвольных функций инструмента нам не известно никакое другое явное представление ядра $AA^{\ast }$.

\section{Собственные значения и функции оператора $A^{\ast } A$}
\label{s:EigsAAstar}

\subsection{Собственные значения и функции оператора $A^{\ast } A$ в одномерном случае для гауссовой функции пучка}
\label{ss:GaussianTool}

Теперь мы опишем более подробно собственные функции и собственные значения оператора $A^{\ast } A$ для $f(x)$, заданной формулой (\ref{eq:Gaussian1}). В этом случае
\begin{eqnarray}
\label{eq:GaussianKernel}
& &
k (x^{\prime } , x^{\prime \prime } ) = \nonumber
\\
& &
\frac{1}{2} \, \frac{1}{\sqrt{4 \pi \sigma ^2 } } \,
e^{-\frac{(x^{\prime } - x^{\prime \prime } )^2 }{4\sigma ^2 } } \biggl[
\mathrm{erf} \left( \frac{x^{\prime }  + x^{\prime \prime }  + 2L}{2 \sigma  } \right) -
\mathrm{erf} \left( \frac{x^{\prime } + x^{\prime \prime } - 2L}{2\sigma } \right) \biggr] .
\end{eqnarray}
Ради простоты мы предполагаем, что $\frac{\sigma }{L} \ll 1$.
Для $x^{\prime }$ и  $x^{\prime \prime }$ близких к нулю, асимптотика $\mathrm{erf} (x)$ при $x \to + \infty$
(см. формулу 7.1.23 в \cite{AbramowitzStegun}) даёт
\begin{eqnarray*}
& &
k (x^{\prime } , x^{\prime \prime } ) \simeq  \frac{1}{\sqrt{4 \pi \sigma ^2 } } \,
e^{-\frac{(x^{\prime } - x^{\prime \prime } )^2 }{4\sigma ^2 } } \biggl[ 1 -
\frac{\sigma \, e^{- \frac{(2L + x^{\prime } + x^{\prime \prime } )^2}{4\sigma ^2}  } }{\sqrt{\pi } (2L + x^{\prime } + x^{\prime \prime })}  -
\frac{\sigma \, e^{- \frac{(2L - x^{\prime } - x^{\prime \prime } )^2}{4\sigma ^2 } } }{\sqrt{\pi } (2L - x^{\prime } - x^{\prime \prime })}  \biggr] \sim
\\
& &
\frac{1}{\sqrt{4 \pi \sigma ^2 } } \,
e^{-\frac{(x^{\prime } - x^{\prime \prime } )^2 }{4\sigma ^2 } } .
\end{eqnarray*}
Из этого выражения мы можем вывести эвристическое описание собственных значений и собственных векторов $A^{\ast } A$. Внутри интервала $[-L,L]$ и вдали от его
границ (точное описание будет дано позже) ядро
$k (x^{\prime } , x^{\prime \prime } )$ асимптотически близко к фундаментальному решению уравнения теплопроводности в момент времени $\tilde{\tau } = \sigma ^2$. Заметим, что пределы интегрирования в интегральном
выражении для ядра (\ref{eq:IntegralForKernel}) могут быть приближенно смоделированы нулевыми граничными 
условиями для уравнения теплопроводности.
Мы ожидаем, что $A^{\ast } A$ будет в некотором роде похож на оператор $e^{ \sigma ^2 \partial_x^2 }$ с нулевыми граничными условиями в точках $-L$ и $L$. Известно, что собственными функциями
$e^{ \tilde{\tau } \partial_x^2 }$ являются
\begin{equation}
\label{eq:HeatEquationEigenfunctions}
\frac{1}{L} \cos \frac{(2n + 1)\pi }{2L} x , \qquad \frac{1}{L} \sin \frac{n \pi }{L} x ,
\end{equation}
с собственными значениями $e^{-\sigma ^2 \frac{(2n + 1)^2\pi ^2}{4L^2}}$ и $e^{-\sigma ^2 \frac{n^2 \pi ^2}{L^2} }$. Здесь мы разделили собственные функции одномерного оператора Лапласа
на чётные (косинусы) и нечётные (синусы). На Рис. 1 мы показываем численно вычисленные первые три собственные функции $A^{\ast } A$ для $\sigma = 2 $ и $L = 80$ вместе с первыми тремя
собственными функциями из (\ref{eq:HeatEquationEigenfunctions}).
\begin{figure}[!tbh]
\centering
\begin{minipage}{0.47\textwidth}
\centering
\includegraphics[width=6.0cm]{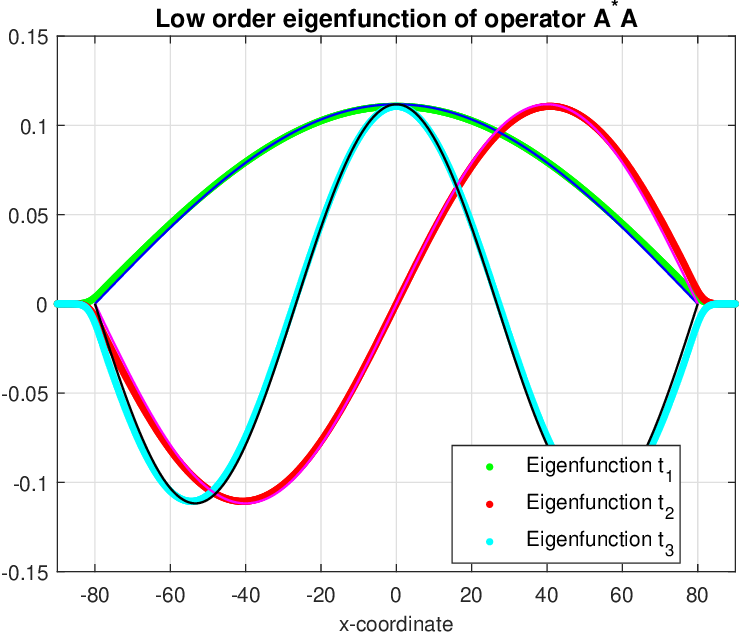}
\end{minipage}
\begin{minipage}{0.47\textwidth}
\centering
\includegraphics[width=6.0cm]{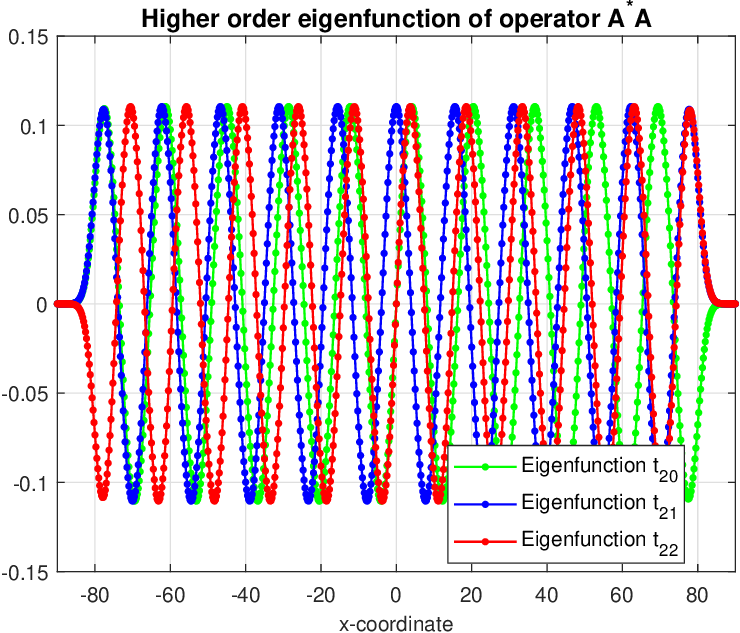}
\end{minipage}
\caption{Собственные функции $A^{\ast } A$ с гауссовым ядром (\ref{eq:GaussianKernel}) показаны на левом рисунке для $\sigma = 2 $ и $L = 80$. Собственные функции низких порядков показаны вместе
с первыми тремя собственными функциями $\partial_x^2 $ с нулевыми граничными условиями (показаны тонкими линиями). На правом рисунке мы показываем некоторые собственные функции более высоких порядков. Все собственные функции были
вычислены с использованием формулы (\ref{eq:discreteKernelAAstar}) для $\Delta $, малого по сравнению с $\sigma $ и $L$.}
\label{fig1}
\end{figure}
Теперь мы покажем, что собственные функции внутри интервала $[-L, L]$ действительно асимптотически близки к функциям косинуса и синуса. Прежде чем сделать это, отметим, что используя неравенство Коши--Буняковского
мы можем показать, что собственные функции (\ref{eq:GaussianKernel}) бесконечно дифференцируемы, более того, аналитичны.

\begin{Prop}
Пусть $x^{\prime } \in [-qL, qL]$ с $q < 1$. Предполагая, что функция $t(x^{\prime \prime } )$ удовлетворяет $t(x^{\prime \prime } ) = e^{ibx^{\prime \prime }}$ для
$x^{\prime \prime } \in [-L, L]$ и $t \in L^1 (\mathbb{R} )$, мы получаем следующее соотношение
\begin{eqnarray}
\label{eq:AsymptoticEigenvalue}
& &
\int_{\mathbb{R} } k (x^{\prime } , x^{\prime \prime } ) t(x^{\prime \prime } ) \, dx^{\prime \prime } =
\\
& &
e^{-b^2\sigma ^2} e^{i b x^{\prime }} \frac{1}{2} \,  \biggl[  \mathrm{erf} \left( \frac{x^{\prime } + L + i b \sigma }{\sqrt{2} \sigma } \right) -
\mathrm{erf} \left( \frac{x^{\prime } - L + i b \sigma }{\sqrt{2} \sigma } \right) \biggr]  +
O \left( \frac{L}{\sigma } \, e^{-\frac{( 1- q)^2 L^2 }{4 \sigma ^2}} \right) . \nonumber
\end{eqnarray}
\end{Prop}
Для доказательства (\ref{eq:AsymptoticEigenvalue}) мы разобьём область интегрирования на три части: $(-\infty , -L)$, $[-L, L]$ и $(L, +\infty )$
и расширим интеграл, содержащий $e^{ibx^{\prime \prime }}$, на всю вещественную ось
\begin{eqnarray}
\label{eq:integralDecompposition}
& &
\int_{\mathbb{R} } k (x^{\prime } , x^{\prime \prime } ) t(x^{\prime \prime } ) \, dx^{\prime \prime } = \int_{-\infty}^{-L} k (x^{\prime } , x^{\prime \prime } ) t(x^{\prime \prime } ) \, dx^{\prime \prime } -
\int_{-\infty}^{-L} k (x^{\prime } , x^{\prime \prime } ) e^{ibx^{\prime \prime }} \, dx^{\prime \prime }  + \nonumber
\\
& &
\int_{\mathbb{R}} k (x^{\prime } , x^{\prime \prime } ) e^{ibx^{\prime \prime }} \, dx^{\prime \prime } -
\int_{L}^{+\infty} k (x^{\prime } , x^{\prime \prime } ) e^{ibx^{\prime \prime }} \, dx^{\prime \prime } +
\int_{L}^{+\infty} k (x^{\prime } , x^{\prime \prime } ) t(x^{\prime \prime } ) \, dx^{\prime \prime } .
\end{eqnarray}
Интеграл по всей вещественной оси может быть явно вычислен с использованием Формулы 13 в Разделе 4.3 из \cite{NgGeller}
\begin{displaymath}
\int_{-\infty}^{+\infty} k_w (x^{\prime } , x^{\prime \prime } ) e^{i b x^{\prime \prime }} \, dx^{\prime \prime } =
e^{-b^2\sigma ^2} e^{i b x^{\prime }} \frac{1}{2} \,  \biggl[  \mathrm{erf} \left( \frac{x^{\prime } + L + i b \sigma }{\sqrt{2} \sigma } \right) -
\mathrm{erf} \left( \frac{x^{\prime } - L + i b \sigma }{\sqrt{2} \sigma } \right) \biggr]   .
\end{displaymath}
Оставшиеся члены могут быть оценены следующим образом
\begin{eqnarray*}
& &
\vert \int_{-\infty}^{-L} k (x^{\prime } , x^{\prime \prime } ) t(x^{\prime \prime } ) \, dx^{\prime \prime } \vert \leq
\int_{-\infty}^{-L} k (x^{\prime } , x^{\prime \prime } ) \vert t(x^{\prime \prime } ) \vert \, dx^{\prime \prime } \leq
\\
& &
\sup_{x^{\prime } \in [-qL, qL] , x^{\prime \prime } \in (-\infty , -L]} \{ k (x^{\prime } , x^{\prime \prime } ) \} \, \Vert t \Vert _{L^1 (\mathbb{R} )} \leq
\frac{1}{\sqrt{4 \pi \sigma ^2 } } \,  e^{-\frac{( 1- q)^2 L^2 }{4 \sigma ^2}} \Vert t \Vert _{L^1 (\mathbb{R} )}
\end{eqnarray*}
и
\begin{eqnarray*}
& &
\vert \int_{-\infty}^{-L} k (x^{\prime } , x^{\prime \prime } ) e^{ibx^{\prime \prime }} \, dx^{\prime \prime } \vert \leq \int_{-\infty}^{-L} k (x^{\prime } , x^{\prime \prime } ) \, dx^{\prime \prime } \leq
\\
& &
\frac{1}{\sqrt{4 \pi \sigma ^2 } } \,  e^{-\frac{( 1- q)^2 L^2 }{4 \sigma ^2}} \int_{\mathbb{R}}  \frac{1}{2} \, \biggl[
\mathrm{erf} \left( \frac{x^{\prime }  + x^{\prime \prime }  + 2L}{2 \sigma  } \right) - \mathrm{erf} \left( \frac{x^{\prime } + x^{\prime \prime } - 2L}{2\sigma } \right) \biggr] \, dx^{\prime \prime } \leq
\\
& &
\frac{1}{\sqrt{\pi } } \frac{2L}{\sigma } \,  e^{-\frac{( 1- q)^2 L^2 }{4 \sigma ^2}}
\end{eqnarray*}
Интегралы по $(L, +\infty )$ оцениваются аналогично.

Из (\ref{eq:AsymptoticEigenvalue}) следует, что волновое число $k_n $ собственной функции $t_n $ и соответствующее
собственное значение связаны соотношением
\begin{equation}
\label{eq:GaussFreqEigenvalue}
\lambda _n = e^{-k_n^2 \sigma^2 } .
\end{equation}
Чтобы показать, насколько хорошо собственные функции аппроксимируются функциями косинуса и синуса внутри $[-L,L]$, мы построили на Рис. 2 графики $t_n (x) /  \cos \sqrt{ - \frac{ \ln \lambda _n }{\sigma^2} } x $ для чётных и
$t_n (x) / \sin \sqrt{ - \frac{ \ln \lambda _n }{\sigma^2} } x$ для нечётных собственных функций для численных значений $\sigma = 2 $ и $L = 80$.
\begin{figure}[!tbh]
\centering
\begin{minipage}{0.47\textwidth}
\centering
\includegraphics[width=6.0cm]{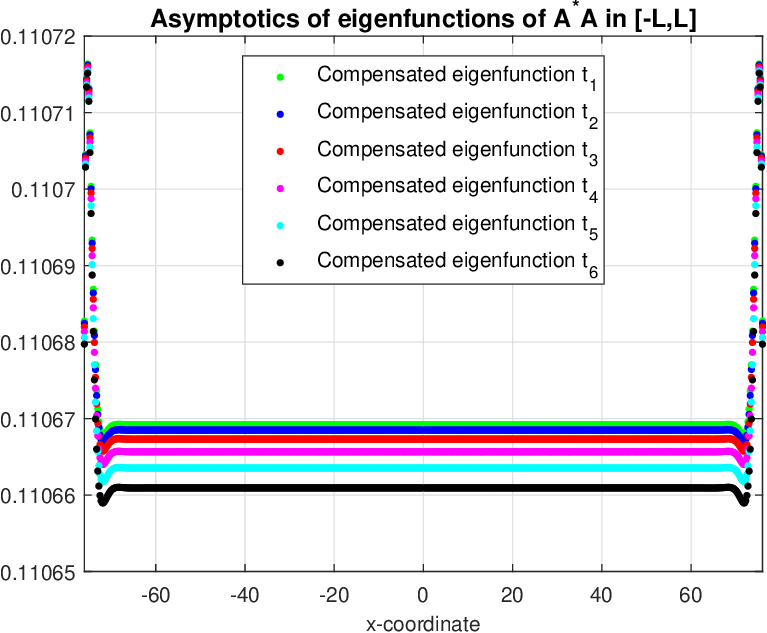}
\end{minipage}
\begin{minipage}{0.47\textwidth}
\centering
\includegraphics[width=6.0cm]{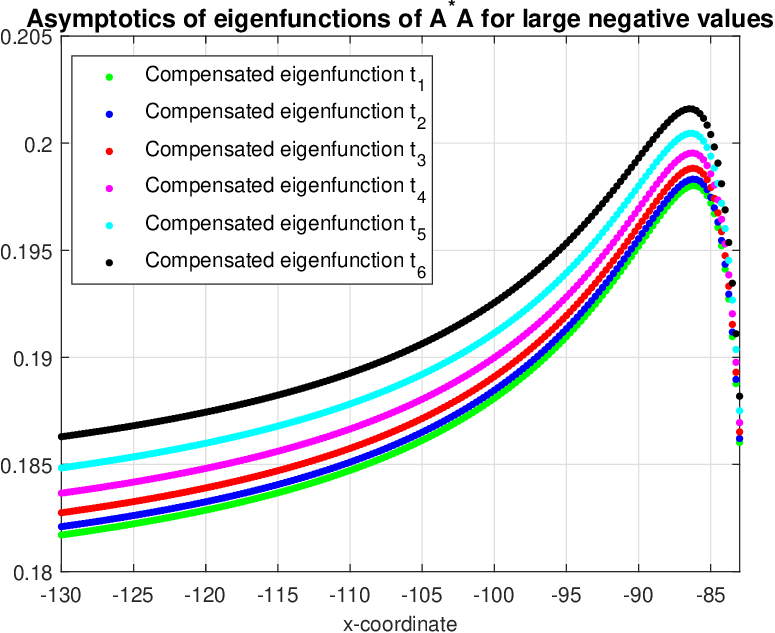}
\end{minipage}
\caption{Собственные функции $t_n (x)$ оператора $A^{\ast } A$ с гауссовым ядром (\ref{eq:GaussianKernel}), делённые на $\cos \sqrt{ - \frac{ \ln \lambda _n }{\sigma^2} } x $ для чётных, соответственно
на $\sin \sqrt{ - \frac{ \ln \lambda _n }{\sigma^2} } x $ для нечётных собственных функций для численных значений $\sigma = 2 $ и $L = 80$, показаны на левом рисунке.
На правом рисунке эти собственные функции скомпенсированы на $(x^{\prime } + L) \, e^{\frac{(x + L)^2}{2\sigma ^2}  } / \sqrt{ - \frac{ \ln \lambda _n }{\sigma^2} }$.}
\label{fig2}
\end{figure}
Заметим, что быстрое убывание собственных значений $\lambda _n $ оператора $A^{\ast } A$ (и, следовательно, сингулярных значений $\sqrt{\lambda _n }$ оператора $A$) с волновым числом 
$k_n$ в уравнении (\ref{eq:GaussFreqEigenvalue}) означает, что эффективность травления, определённая в уравнении (\ref{eq:singVectors}), для сильно осциллирующих карт времени экспозиции
также очень быстро убывает с ростом волнового числа. Другими словами, коррекция мелкомасштабных структур в профиле поверхности требует больших амплитуд времени экспозиции.

Анализ численного моделирования собственных функций \cite{Patent} показывает, что в первом приближении для малых значений $\frac{\sigma }{L}$ волновые числа низкого порядка могут быть описаны выражением
\begin{displaymath}
k_n = \frac{\pi n}{2L} \left( 1 - \frac{\pi }{4} \frac{\sigma }{L} + ... \right) .
\end{displaymath}
К сожалению, нам не удалось определить асимптотику $t_n (x)$ при $x \to -\infty $ или $x \to + \infty $. Формула (\ref{eq:AsymptoticEigenvalue}) даёт намёк на возможную скорость убывания
$t_n (x)$ вида
\begin{eqnarray*}
& &
t_n (x) \sim - \frac{1}{x + L} \, e^{- \frac{(x + L)^2}{2\sigma ^2}  } , \qquad x \to -\infty  ,
\\
& &
t_n (x) \sim \frac{1}{x - L} \, e^{- \frac{(x - L)^2}{2\sigma ^2}  } , \qquad x \to +\infty  .
\end{eqnarray*}
Более подробный анализ требует дополнительных исследований, которые выходят за рамки данной работы.

Стоит отметить, что собственные функции $A^{\ast } A$ представляют собой своего рода локализованные функции косинуса и синуса с быстрым убыванием на (вещественной) бесконечности, которые являются аналитическими.

\subsection{Некоторые замечания о собственных значениях и собственных функциях $A^{\ast } A$ в общем одномерном случае}
\label{ss:DiscreteCase}

Мы ожидаем, что для произвольных функций инструмента, при некоторых предположениях относительно $f(x)$, асимптотика 
собственных функций с не равным нулю $\lambda _n $ внутри
интервала $[-L,L]$ также будет иметь вид $e^{ikx}$ при $L \to \infty $.  Действительно, для абсолютно интегрируемой функции 
$t$, равной $e^{ikx}$ при $x \in [-L, L]$,
мы можем использовать разложение (\ref{eq:integralDecompposition}). Функции инструмента должны быть такими, чтобы интегралы
\begin{eqnarray*}
& &
\vert \int_{-\infty}^{-L} k (x^{\prime } , x^{\prime \prime } ) t(x^{\prime \prime } ) \, dx^{\prime \prime } \vert \leq
\sup_{x^{\prime } \in [-qL, qL] , x^{\prime \prime } \in (-\infty , -L]} \{ k (x^{\prime } , x^{\prime \prime } ) \} \, \Vert t \Vert _{L^1 (\mathbb{R} )} ,
\\
& &
\vert \int_{L}^{\infty} k (x^{\prime } , x^{\prime \prime } ) t(x^{\prime \prime } ) \, dx^{\prime \prime } \vert \leq
\sup_{x^{\prime } \in [-qL, qL] , x^{\prime \prime } \in [L, +\infty )} \{ k (x^{\prime } , x^{\prime \prime } ) \} \, \Vert t \Vert _{L^1 (\mathbb{R} )} ,
\end{eqnarray*}
и
\begin{eqnarray*}
& &
\vert \int_{-\infty}^{-L} k (x^{\prime } , x^{\prime \prime } ) e^{ibx^{\prime \prime }} \, dx^{\prime \prime } \vert \leq
\int_{-L}^{L} f(x - x^{\prime }) \left( \int_{-\infty}^{-L} f(x - x^{\prime \prime } ) \, dx^{\prime \prime } \right) \, dx ,
\\
& &
\vert \int_{L}^{\infty} k (x^{\prime } , x^{\prime \prime } ) e^{ibx^{\prime \prime }} \, dx^{\prime \prime } \vert \leq
\int_{-L}^{L} f(x - x^{\prime }) \left( \int_{L}^{\infty} f(x - x^{\prime \prime } ) \, dx^{\prime \prime } \right) \, dx ,
\end{eqnarray*}
были асимптотически малыми при $L \to \infty $. Тогда мы остаёмся с вкладом
\begin{eqnarray*}
& &
\int_{\mathbb{R}} k (x^{\prime } , x^{\prime \prime } ) e^{ibx^{\prime \prime }} \, dx^{\prime \prime } \sim \sqrt{2 \pi } \left( \mathcal{F} f \right) (k) e^{ik x^{\prime }} \int_{x^{\prime } - L}^{x^{\prime } + L}
f(-x)  e^{-ik x} \, dx \sim
\\
& &
2 \pi  \left( \mathcal{F} f \right) (k) \left( \overline{\mathcal{F} f} \right) (k) \, e^{ik x^{\prime }} ,
\end{eqnarray*}
где преобразование Фурье $( \mathcal{F} f) (k)$ функции $f(x)$ определяется как
\begin{displaymath}
( \mathcal{F} f) (k) = \frac{1}{\sqrt{2 \pi }} \int_{\mathbb{R}} f(x)  e^{-ik x} \, dx .
\end{displaymath}
Таким образом, собственные значения $A^{\ast } A$ в общем случае асимптотически равны $2 \pi  \left( \mathcal{F} f \right) (k)
\left( \overline{\mathcal{F} f} \right) (k)$.
Таким образом, скорость убывания собственных значений с $k$ зависит от $\left( \mathcal{F} f \right) (k)$. Заметим, что для
функций инструмента с немонотонным убыванием $\vert \left( \mathcal{F} f \right) (k) \vert ^2 $, собственные функции упорядочены по
величине собственных значений, а не по величине соответствующего волнового вектора $k$.

\begin{figure}[!tbh]
\centering
\begin{minipage}{0.47\textwidth}
\centering
\includegraphics[width=6.0cm]{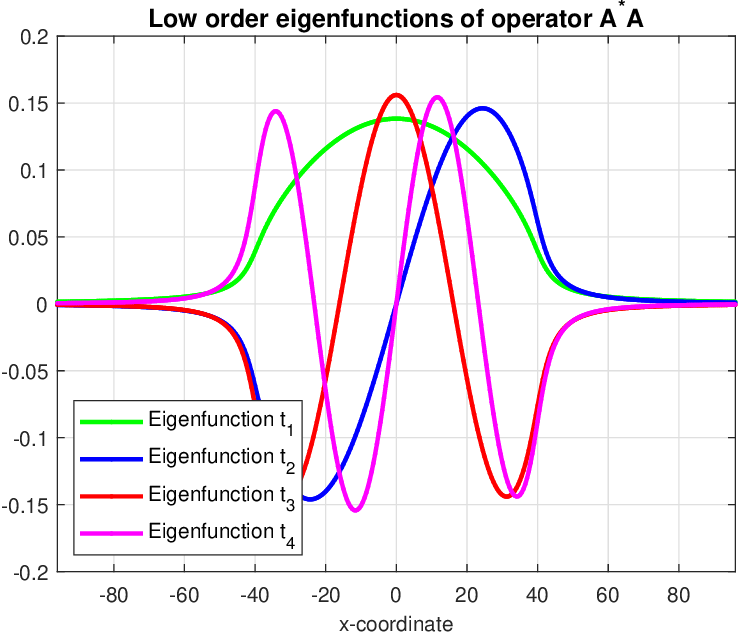}
\end{minipage}
\begin{minipage}{0.47\textwidth}
\centering
\includegraphics[width=6.0cm]{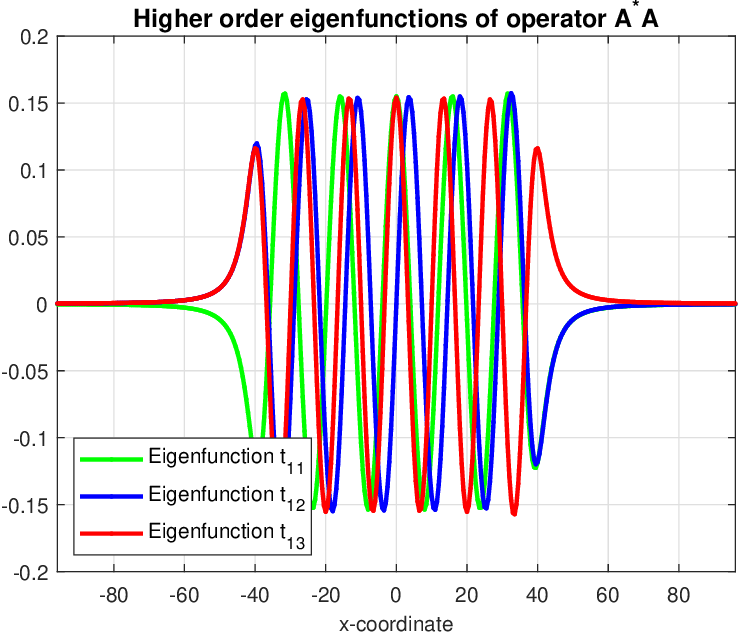}
\end{minipage}
\caption{Собственные функции $A^{\ast } A$ с  одномерным пуассоновым ядром (распределение Коши) (\ref{eq:GaussianKernel}) показаны на левом рисунке для $\sigma = 2 $ и $L = 40$. 
Все собственные функции были
вычислены с использованием формулы (\ref{eq:discreteKernelAAstar}) для $\Delta $, малого по сравнению с $\sigma $ и $L$.}
\label{fig14}
\end{figure}

Для оценки справедливости сделанных выше оценок были численно рассчитаны собственные функции в нескольких случаях, например, для функции
инструмента, заданной с помощью одномерного ядра Пуассона (распределения Коши), см. Рис. \ref{fig14}.
Из результатов численного моделирования видно, что хотя поведение собственных функций внутри интервала $[-L,L]$ аналогично случаю гауссова ядра, 
размеры переходной области вокруг границ интервала значительно больше. Это объясняется тем, что скорость убывания ядра пуассона гораздо медленнее,
чем у гауссова ядра и, таким образом, граничные эффекты играют значительно большую роль.

\subsection{Собственные числа и функции оператора $A^{\ast } A$ с гауссовым ядром на диске}

В двумерном случае явное представление ядра оператора $A^{\ast } A$, аналогичное (\ref{eq:GaussianKernel}), может быть получено только для простых
областей, таких как прямоугольники или диски, и гауссовых функций влияния инструмента. Для прямоугольных областей собственные функции могут быть построены как тензорные произведения
одномерных собственных функций. На диске, используя полярные координаты, собственные функции могут быть разложены в соответствии с их симметриями относительно группы
вращений $O(2)$ на секторы $\cos m \theta $, $m = 0, 1, 2,...$ и $\sin m\theta $, $m = 1, 2,...$. Как мы увидим ниже, для каждого $m$ ядро может быть записано в виде интеграла, включающего
модифицированные функции Бесселя $I_m $. По аналогии с одномерным случаем, для больших
радиусов диска собственные функции асимптотически близки к собственным функциям оператора $e^{\sigma ^2 \Delta }$ с нулевыми граничными условиями. Аналогичные соображения
применимы и для более высоких размерностей, где для разложения собственных функций по секторам сферических функций нужно использовать сферические гармоники.

Предположим, что область измерений является диском с радиусом $R$. Тогда ядро оператора $A^{\ast } A$ задано как
\begin{eqnarray*}
& & k (x_1^{\prime } , x_2^{\prime } , x_1^{\prime \prime } , x_2^{\prime \prime } ) =
\\
& & \int_0^R \int_0^{2\pi }  f_{\sigma } (r \cos \theta  - x_1^{\prime } , r \sin \theta - x_2^{\prime } )  
f_{\sigma } (r \cos \theta - x_1^{\prime \prime } , r \sin \theta - x_2^{\prime \prime } )  \, r dr d\theta  .
\end{eqnarray*}
Переходя к полярным координатам $ (r^{\prime } , \theta ^{\prime }) $ и $ (r^{\prime \prime } , \theta ^{\prime \prime })$ для переменных $(x^{\prime }_1 , x^{\prime }_2 )$ и
$(x^{\prime \prime }_1 , x^{\prime \prime }_2 )$, мы получим выражение 
\begin{eqnarray*}
& & k (r^{\prime } , r^{\prime \prime } , \phi ) = 
\\
& & \frac{1}{(2\pi \sigma ^2 )^2 } \, e^{- \frac{r^{\prime \, 2} }{2\sigma ^2 } } e^{- \frac{r^{\prime \prime \, 2} }{2\sigma ^2 } }
\int_0^R  e^{- \frac{r^2 }{\sigma ^2 } } \int_0^{2 \pi }  \exp \Bigl[ \frac{r}{\sigma ^2 }
( r^{\prime } \cos \theta + r^{\prime \prime } \cos (\theta + \phi ) )  \Bigr] \, r dr d\theta  ,
\end{eqnarray*}
где $\phi = \theta ^{\prime } - \theta ^{\prime \prime }$.
Прежде всего, заметим, что $ k (r^{\prime } , r^{\prime \prime } , \phi ) $ и интеграл
\begin{equation*}
I (r, r^{\prime } , r^{\prime \prime } , \phi ) = \int_0^{2 \pi }  \exp \Bigl[ \frac{r}{\sigma ^2 }
( r^{\prime } \cos \theta + r^{\prime \prime } \cos (\theta + \phi ) )  \Bigr] \, d\theta  ,
\end{equation*}
являются чётными функциями по $\phi $. Разлагая интеграл $I (r, r^{\prime } , r^{\prime \prime } , \phi )$ в ряд Фурье по $\phi$ 
мы можем вычислить его коэффициенты как
\begin{eqnarray*}
& & a_n  (r, r^{\prime } , r^{\prime \prime } ) = \frac{1}{\pi } \int_0^{2\pi } I (r, r^{\prime } , r^{\prime \prime } , \phi )
\cos n \phi \, d\phi = 
\\
& & \frac{1}{\pi } \int_0^{2 \pi } e^{\frac{r r^{\prime } }{\sigma ^2 } \cos \theta } \cos n \theta \, d\theta \, \int_0^{2 \pi }
e^{ \frac{r r^{\prime \prime } }{\sigma ^2 }  \cos \phi   } \cos n \phi   \,  d\phi ,
\end{eqnarray*}
и, таким образом, используя формулу (9.6.20) из \cite{AbramowitzStegun}
\begin{equation}
I_n (z) = \frac{1}{\pi  } \int_0^{\pi } e^{z \cos \theta } \cos n \theta \, d\theta ,
\end{equation}
мы получаем следующее выражение
\begin{equation*}
a_n  (r, r^{\prime } , r^{\prime \prime } ) =
4 \pi \, I_n \left( \frac{r r^{\prime } }{\sigma ^2 } \right) I_n \left( \frac{r r^{\prime \prime } }{\sigma ^2 } \right)  .
\end{equation*}
Следовательно,  ядро $k (r^{\prime } , r^{\prime \prime } , \phi )  $ оператора $A^{\ast } A$ может быть 
представлено в виде следующего ряда
\begin{eqnarray}
\label{eq:ModBesselKernelSeries}
& & k (r^{\prime } , r^{\prime \prime } , \phi ) = \nonumber
\\
& & \frac{1}{\pi \sigma ^4 } \,
e^{- \frac{r^{\prime \, 2} }{2\sigma ^2 } } e^{- \frac{r^{\prime \prime \, 2} }{2\sigma ^2 } }
\sum_{n=0}^{\infty } \cos n \phi \,  \int_0^R  e^{- \frac{r^2 }{\sigma ^2 } }
I_n \left( \frac{r r^{\prime } }{\sigma ^2 } \right) I_n \left( \frac{r r^{\prime \prime } }{\sigma ^2 } \right)  \, r dr  ,
\end{eqnarray}
Для удобства введем обозначения
\begin{equation*}
k^{(n)}  (r^{\prime } , r^{\prime \prime } ) =
\frac{1}{\pi \sigma ^4 } \,
e^{- \frac{r^{\prime \, 2} }{2\sigma ^2 } } e^{- \frac{r^{\prime \prime \, 2} }{2\sigma ^2 } } \,  \int_0^R  e^{- \frac{r^2 }{\sigma ^2 } }
I_n \left( \frac{r r^{\prime } }{\sigma ^2 } \right) I_n \left( \frac{r r^{\prime \prime } }{\sigma ^2 } \right)  \, r dr  .
\end{equation*}

Чтобы иметь возможность использовать соображения симметрии по отношению ко вращениям, 
представим карту травления как
\begin{displaymath}
t (r^{\prime \prime } , \theta ^{\prime \prime } ) = \sum_{n=0}^{\infty }
t^{(n)}_{c} (r^{\prime \prime } ) \cos n \theta ^{\prime \prime } +
f^{(n)}_{s} (r^{\prime \prime } ) \sin n \theta ^{\prime \prime } .
\end{displaymath}
Тогда действие оператора $A^{\ast } A$ на карту травления $t (r, \theta )$ выражается как
\begin{eqnarray*}
& & (A^{\ast } A t) (r^{\prime } , \theta ^{\prime } ) = 
\pi \sum_{n=0}^{\infty }  \cos n \theta ^{\prime }  \,   \int_0^{\infty }
k^{(n)}  (r^{\prime } , r^{\prime \prime } )
\, t^{(n)}_{c} (r^{\prime \prime } )  \, r^{\prime \prime } dr^{\prime \prime }  +
\\
& & \pi \sum_{n=1}^{\infty } \sin n \theta ^{\prime } \, \int_0^{\infty }
k^{(n)}  (r^{\prime } , r^{\prime \prime } )
\, t^{(n)}_{s} (r^{\prime \prime } ) \, r^{\prime \prime } dr^{\prime \prime }  .
\end{eqnarray*}
Из этого представления мы видим, что оператор $A^{\ast } A$ сохраняет угловую гармонику и чётность,
так что достаточно изучить действие оператора на угловые секторы $\cos n \theta $ и $\sin n \theta $, 
рассматривая раздельно интегральные операторы, определенные ядрами
$k^{(n)}  (r, r^{\prime } )$ на радиальных компонентах карт травления $t^{(n)}_{c} (r)$ и $t^{(n)}_{s} (r)$.

\subsection{Собственные значения и функции оператора $A^{\ast } A$ в общем случае}

Что происходит для областей измерений, геометрия которых не ограничивается простейшими случаями, такими как диск или прямоугольник? Мы ожидаем, что для функций инструмента,
размеры которых асимптотически малы по сравнению с размером области, собственные функции $A^{\ast } A$ близки к собственным функциям оператора $e^{\sigma ^2 \Delta }$.
По построению, собственные функции оператора $e^{\sigma ^2 \Delta }$ с нулевыми граничными условиями идентичны собственным функциям оператора Лапласа $\Delta $
с нулевыми граничными условиями, которые хорошо известны как собственные состояния бесконечно глубокой квантовой ямы или квантового бильярда на $\Omega $. Следовательно, мы ожидаем,
что собственные функции $A^{\ast } A$ будут демонстрировать явления, подобные наблюдаемым в квантовых бильярдах.

\begin{figure}[!tbh]
\centering
\includegraphics[width=10.0cm]{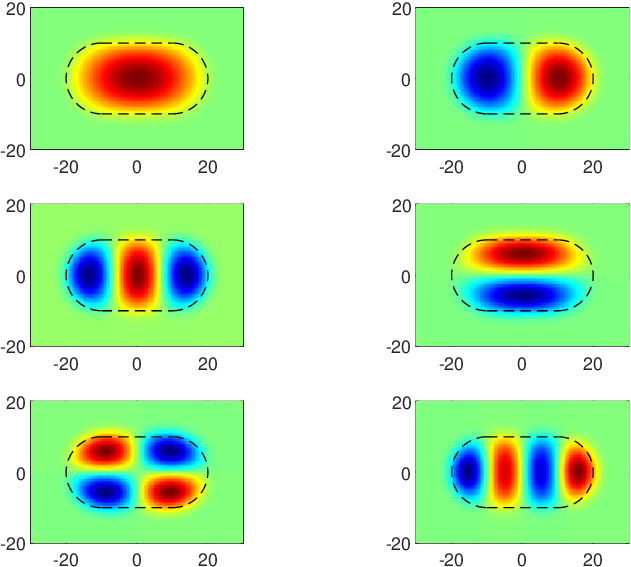}
\caption{Первые шесть собственных функций оператора $A^{\ast } A$ с функцией инструмента, заданной гауссовым ядром (\ref{eq:GaussianKernel}) с $\sigma = 2 $. Область измерений
представляет собой стадион Бунимовича шириной $10$. Контур стадиона показан пунктирной линией.}
\label{fig5}
\end{figure}

\begin{figure}[!tbh]
\centering
\includegraphics[width=10.0cm]{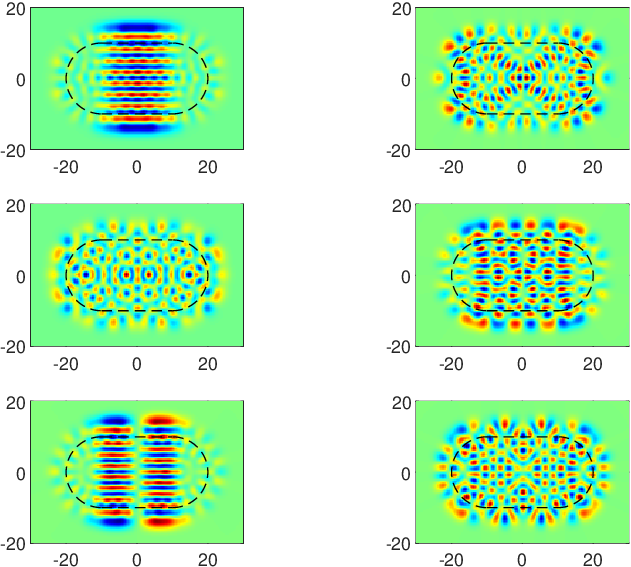}
\caption{Собственные функции высокого порядка оператора $A^{\ast } A$ с функцией инструмента, заданной гауссовым распределением с $\sigma = 2 $. Область измерений
представляет собой стадион Бунимовича шириной $10$. Контур стадиона показан пунктирной линией. Заметим, что собственные функции на верхнем левом и нижнем левом рисунках локализованы
только на части области измерений $\Omega$.}
\label{fig6}
\end{figure}

В качестве примера мы вычислили собственные функции оператора $A^{\ast } A$ в случае, когда $\Omega $ представляет собой стадион Бунимовича, см. \cite{Gutzwiller}. На Рис. \ref{fig5} показаны 
первые шесть собственных функций, упорядоченных по величине их собственных значений. В соответствии с анализом в Разделе \ref{ss:PropertiesAAStar}, первая собственная функция может быть выбрана
положительной. Следующие собственные функции показывают некоторые крупномасштабные колебания, подобно тому, как это имеет место для таких областей, как диск или прямоугольник.
Картина становится более интересной для собственных функций высокого порядка, показанных на Рис. \ref{fig6}. Как видно, некоторые собственные функции распределены более или менее 
равномерно по всей области, тогда как другие локализованы на определённых подмножествах области измерений.
Это явление, в особенности в связи с квантовым хаосом, безусловно заслуживает дальнейшего изучения.

Отметим, что аналогично одномерному случаю, для произвольных функций инструмента, мы ожидаем, что собственные функции $A^{\ast } A$ асимптотически близки к
$e^{i \langle {\bm k} , {\bm x} \rangle _{\mathbb{R}^n }}$ внутри области $\Omega $, с собственными значениями, приблизительно задаваемыми выражением
$(2\pi )^n \vert \left( \mathcal{F} f \right) ({\bm k}) \vert ^2$.

\section{Численный пример} 
\label{s:Practice}

До сих пор наше обсуждение оставалось на довольно абстрактном уровне. Чтобы оценить потенциальную применимость подхода, основанного на формализме, представленном в разделе \ref{ss:InfiniteDimLinEq},
мы представим численный пример, в котором покажем, как вычислить карту травления по заданному набору данных измерений. При этом мы намеренно пренебрегаем некоторыми ограничениями,
возникающими в практических приложениях, чтобы излишне не усложнять изложение. Затем мы обсудим преимущества и недостатки использованного метода.

\begin{figure}[!tbh]
\begin{minipage}{1.0\textwidth}
\centering
\includegraphics[trim={0 3.5cm 0 3.4cm}, clip, width=14.0cm]{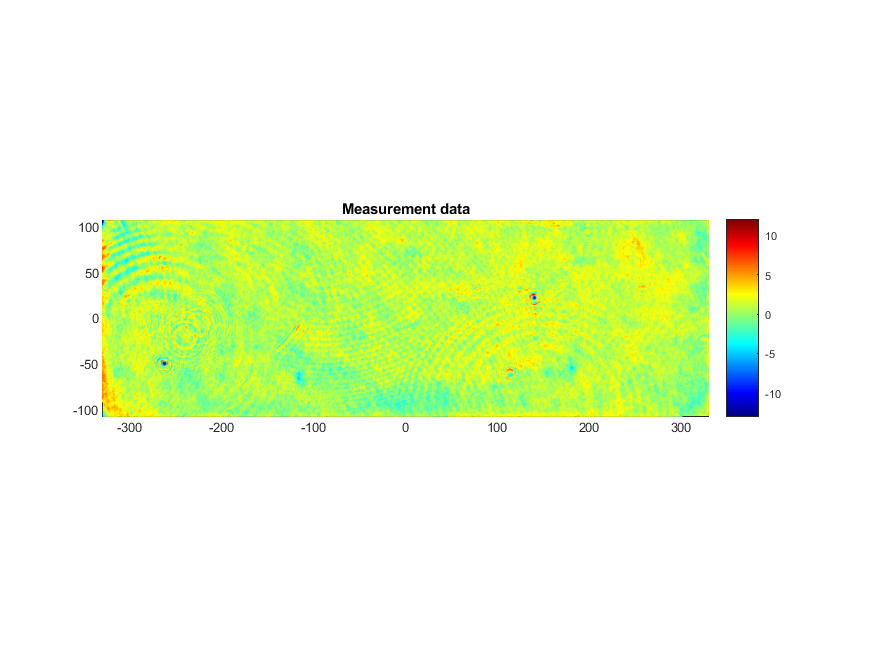}
\end{minipage}
\begin{minipage}{1.0\textwidth}
\centering
\includegraphics[trim={0 3.5cm 0 3.4cm}, clip, width=14.0cm]{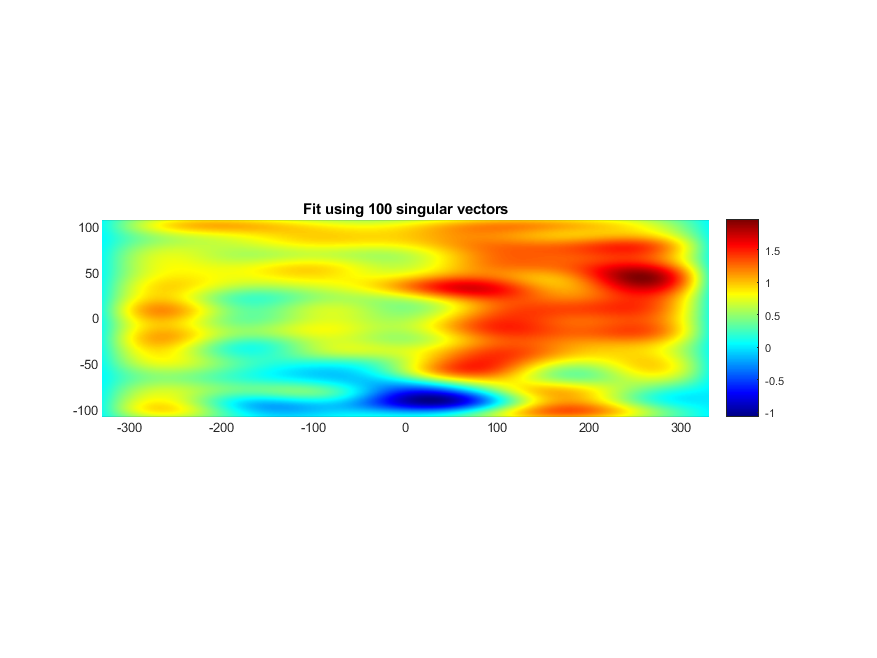} 
\end{minipage}
\begin{minipage}{1.0\textwidth}
\centering
\includegraphics[trim={0 3.5cm 0 3.4cm}, clip, width=14.0cm]{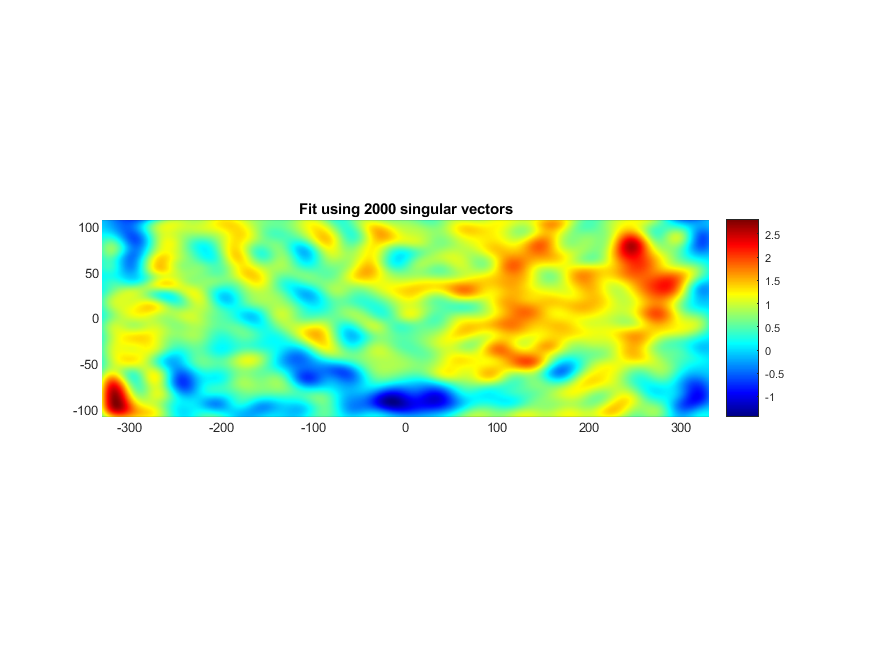} 
\end{minipage}
\caption{Пример результатов измерений, заданных на прямоугольной области, показан на верхнем рисунке. На среднем рисунке показаны восстановленные данные измерений, полученные
с помощью точечного метода наименьших квадратов с использованием $100$ сингулярных векторов. На нижнем рисунке показаны восстановленные данные измерений, полученные
с помощью точечного метода наименьших квадратов с использованием $2000$ сингулярных векторов.}
\label{fig7}
\end{figure}

Данные измерений $h (x_1, x_2 )$, который мы используем, заданы на дискретной сетке в прямоугольной области и показаны на Рис. \ref{fig7}.
Размер области измерений в этом случае составляет $330 \times 107$ пикселей. Для построения карты травления мы применяем усечённую версию формулы (\ref{eq:pseudoinverse}), то есть
обрываем бесконечные суммы в указанной формуле на некотором номере усечения $N_{tr}$.

Функция влияния инструмента в нашем примере задаётся двумерным центрированным гауссовым распределением с $\sigma = 6$ пикселей.
Как отмечено в предыдущем разделе, собственные функции оператора $A^{\ast } A$
могут быть построены как тензорные произведения $t_k (x_1 ) \otimes t_l (x_2 ) $ соответствующих одномерных собственных функций $t_k (x_1 ) $, $t_l (x_2 )$ и образуют 
характерный знакопеременный узор типа шахматная доска. Собственные значения $\lambda _{kl}$ являются произведениями собственных значений $\lambda _k $ и $\lambda _l $, 
соответствующих собственным функциям $t_k $ и $t_l$.
Для вычислительных целей двумерная последовательность собственных значений $\lambda _{kl}$ упорядочивается по величине, порождая таким образом одномерную последовательность $\lambda _n $.
Полученное таким образом упорядочение также используется для упорядочивания собственных функций $t_k (x_1 ) \otimes t_l (x_2 ) $, задавая последовательность собственных функций $t_n (x_1 , x_2 )$.
Эта последовательность собственных функций обрывается на $N_{tr}$ и состоит из функций $\{ t_n (x_1 , x_2 ) ,\ldots , t_{N_{tr}} (x_1 , x_2 ) \}$. Наряду с собственными функциями
$t_n (x_1 , x_2 ) $ мы вычисляем левые сингулярные векторы $h_n (x_1 , x_2 ) $ оператора $A$, определённые в формуле (\ref{eq:LeftSingVec}).

\begin{figure}[!tbh]
\begin{minipage}{1.0\textwidth}
\centering
\includegraphics[trim={0 3.5cm 0 3.4cm}, clip, width=14.0cm]{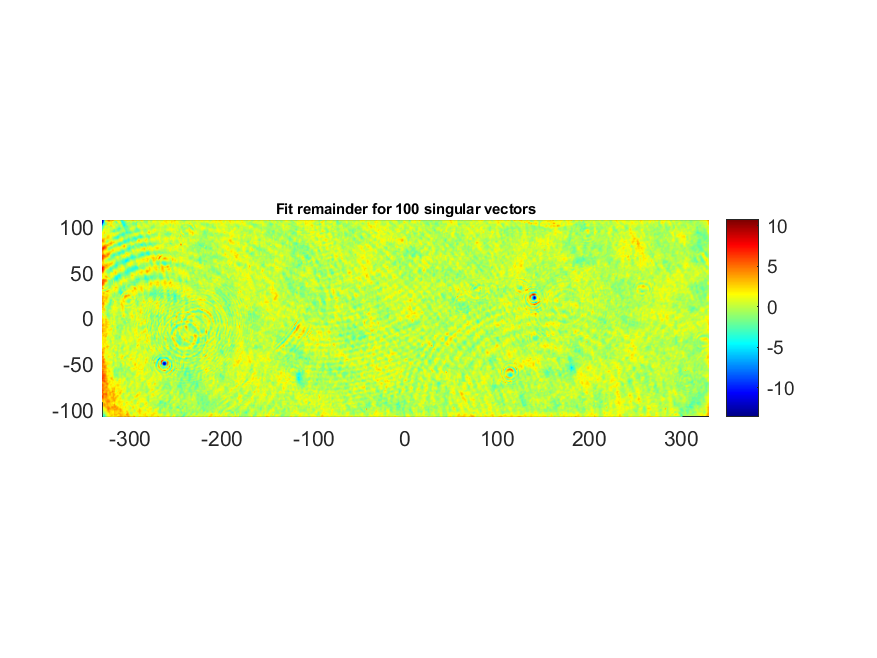}
\end{minipage}
\begin{minipage}{1.0\textwidth}
\centering
\includegraphics[trim={0 3.5cm 0 3.4cm}, clip, width=14.0cm]{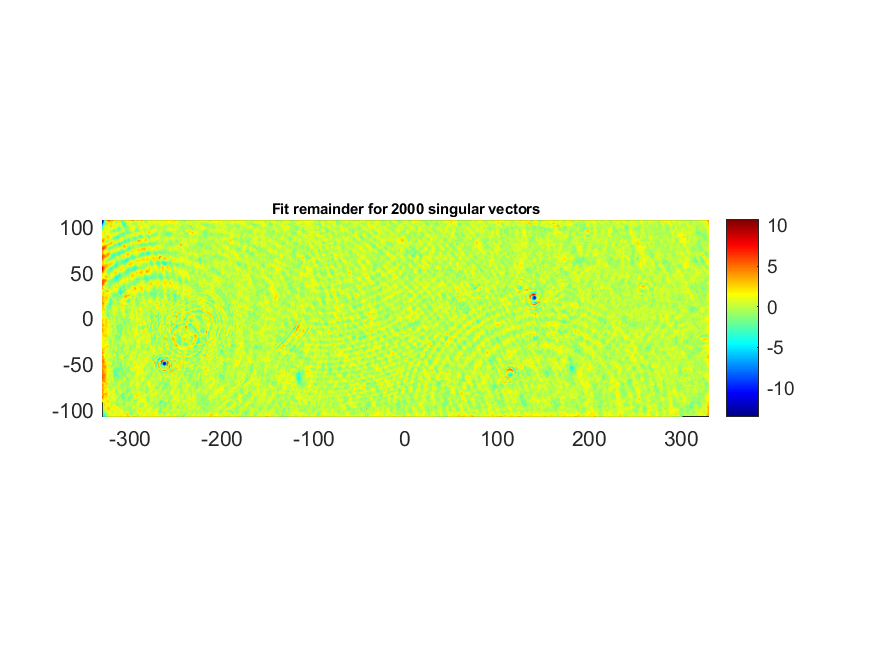} 
\end{minipage}
\caption{Расхождение между данными измерений и подогнанными (восстановленными) данными. На верхнем рисунке показано расхождение при использовании $100$ сингулярных 
векторов, тогда как нижний
соответствует $2000$ сингулярным векторам.}
\label{fig8}
\end{figure}

\begin{figure}[!tbh]
\begin{minipage}{1.0\textwidth}
\centering
\includegraphics[trim={0 3.5cm 0 3.4cm}, clip, width=14.0cm]{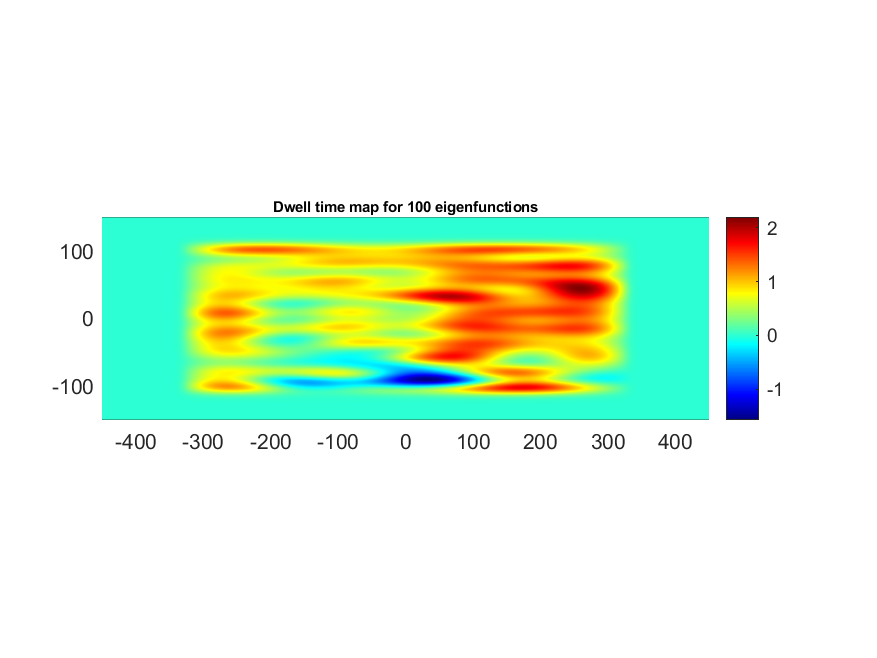}
\end{minipage}
\begin{minipage}{1.0\textwidth}
\centering
\includegraphics[trim={0 3.5cm 0 3.4cm}, clip, width=14.0cm]{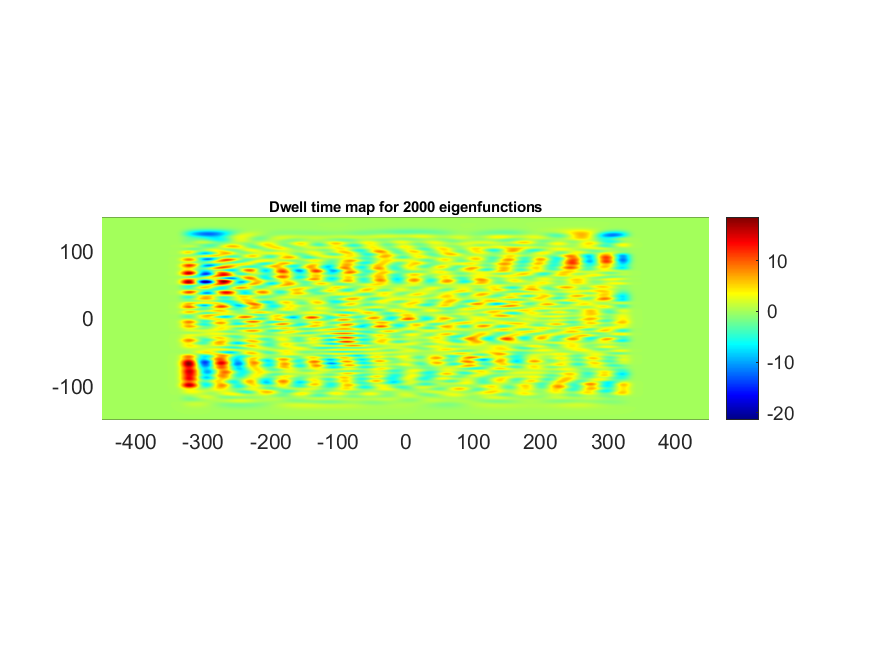} 
\end{minipage}
\caption{Карты травления, вычисленные для $100$ (верхний рисунок) и $2000$ (нижний рисунок) сингулярных векторов. 
Стоит отметить значительно более высокие амплитуды времён экспозиции в случае, когда
используются $2000$ сингулярных векторов.}
\label{fig9}
\end{figure}

На следующем шаге мы вычисляем коэффициенты разложения $h (x_1, x_2 )$ по последовательности сингулярных векторов $\{ h_n (x_1 , x_2 ) ,\ldots , h_{N_{tr}} (x_1 , x_2 ) \}$.
В силу причин, которые будут обсуждены в следующих разделах (для вычислительных деталей см. раздел \ref{s:Filtering}, в частности формулу (\ref{eq:EmpiricalErrorMeas}), где в нашем случае
$\gamma = 0$), это разложение выполняется с помощью поточечной подгонки $h (x_1, x_2 )$ 
методом наименьших квадратов к линейной комбинации $h_n (x_1, x_2 )$.
Полученные результаты подгонки показаны на Рис. \ref{fig7} для двух порядков усечения: (i) для $N_{tr} = 100$ и (ii) $N_{tr} = 2000$. Расхождение между исходными и восстановленными данными измерений
показано на Рис. \ref{fig8}. Заметим, что восстановленные данные измерений можно рассматривать как отфильтрованные данные, поскольку они гораздо глаже, чем исходные, см. раздел \ref{s:Filtering}.
Заметим, что для $N_{tr} = 2000$ длина усечения согласно формуле (\ref{eq:LengthScale}) в разделе \ref{s:Filtering} составляет примерно $16$ пикселей.
Когда коэффициенты разложения известны, соответствующие карты травления, показанные на Рис. \ref{fig9}, легко вычислить из линейной комбинации
собственных функций $t_n (x_1, x_2 )$.

Вычисление карт травления было выполнено с использованием нескольких порядков усечения, от $N_{tr} = 100$ до $N_{tr} = 4000$. Наилучшие результаты подгонки были
получены для $N_{tr} = 2000$ ценой значительного роста времён экспозиции.
Для $N_{tr} > 2000$ результаты подгонки начинают расходиться из-за присущей данной задаче неустойчивости, обсуждавшейся в разделе \ref{s:FuncAna}. Эту неустойчивость подгонки
можно подавить, например, используя регуляризацию, вводя дополнительный параметр, такой, как использованный в разделе \ref{s:Filtering} в формуле (\ref{eq:EmpiricalErrorMeas}).

Что касается практического вычисления карт травления, преимущество метода, который мы использовали в этом разделе, заключается в его простоте и
в том, что все его шаги хорошо определены в математическом смысле. Однако, чтобы сделать его применимым в реальных приложениях, необходимо учесть несколько требований.
Прежде всего, должно соблюдаться ограничение положительности времени экспозиции. Наиболее очевидный способ сделать это — сформулировать
задачу вычисления карты травления как задачу условной оптимизации с квадратичной целевой функцией типа (\ref{eq:QuadMeritFunction}) и ограничением положительности,
наложенным на времена экспозиции. Во-вторых, в зависимости от типа механизма, с помощью которого ионный пучок перемещается по корректируемой поверхности, могут
возникать некоторые динамические ограничения. В-третьих, на некотором уровне точности точная форма траектории, вдоль которой перемещается пучок, начинает иметь значение и,
следовательно, должна быть принята во внимание.

\section{Задача вычисления карт травления в формализме гильбертовых пространств с воспроизводящим ядром и 
радиальных базисных функций} 
\label{s:RKHS}

Вычислительный подход к получению восстанавливающих функций $t$, основанный на соотношении (\ref{eq:Ax=b}) в рамках гильбертовых пространств, имеет один серьёзный недостаток,
который существенно затрудняет его использование в 
практических приложениях, а именно: знак равенства в (\ref{eq:Ax=b}) понимается в смысле $L^2$. В приложениях, однако, требуется поточечная сходимость.
Обычно, имея результат измерений $h({\bm x})$ в виде конечномерной матрицы,
пытаются получить восстанавливающие функции посредством их разложения по базису собственных функций $t_n ({\bm x})$, см. (\ref{eq:pseudoinverse}), причём количество
собственных векторов конечно и, следовательно, должно быть усечено до разумного предела. Вычисление скалярных произведений $ \langle A^{\ast } t_n , h \rangle$ — это одна часть проблемы,
поскольку о свойствах сходимости соответствующих интегралов известно немного. Гораздо более серьёзной проблемой является то, что сходимость самого разложения (\ref{eq:pseudoinverse})
оказывается довольно плохой, если рассматривать её поточечно. На практике обычно аппроксимируют конечную взвешенную сумму левых сингулярных векторов $h_n$ поточечно
к данным измерений, см. раздел \ref{s:Filtering}.

\subsection{Непрерывный случай и гильбертовы пространства с воспроизводящими ядрами}

Мы можем рассмотреть вопросы поточечной аппроксимации
в рамках гильбертовых пространств с воспроизводящим ядром (RKHS), см. \cite{Garcia}.
Мы строим наше гильбертово пространство с воспроизводящим ядром как подпространство пространства результатов измерений $L^2 \left( \Omega , \lambda \right)$ с помощью оператора
$A A^{\ast } : L^2 \left( \Omega , \lambda \right) \longrightarrow L^2 \left( \Omega , \lambda \right)$ с ядром оператора, заданным как
\begin{displaymath}
k ({\bm x}^{\prime } , {\bm x}^{\prime \prime }) = (f \ast f)({\bm x}^{\prime } - {\bm x}^{\prime \prime }) ,
\end{displaymath}
см. уравнение (\ref{eq:dualOperator}).
В рамках ранее сделанных предположений этот оператор компактен и, следовательно, его обратный оператор $(AA^{\ast })^{-1}$ является строго монотонным, так что мы
можем определить наше гильбертово пространство с воспроизводящим ядром
как энергетическое\footnote{Энергетическое пространство $\mathcal{H}_{(AA^{\ast })^{-1}}$ не следует путать с 
пространством энергии, введённым, например, в \cite{BahChemD}.}
пространство $\mathcal{H}_{(AA^{\ast })^{-1}}$ оператора $(AA^{\ast })^{-1}$, снабжённое скалярным произведением
\begin{displaymath}
\langle g , h \rangle _{\mathcal{H}_{(AA^{\ast })^{-1}}} = \langle (AA^{\ast })^{-1} g , h \rangle _{L^2 \left( \Omega , \lambda \right)} ,
\end{displaymath}
см. \cite{Zeidler}.
Воспроизводящее ядро тогда в точности есть $(f \ast f)({\bm x} - {\bm x}^{\prime } )$. Пусть ${\bm x} _i $, $i = 1,...,N$ будут точками дискретизации, т.е. точками, в которых результат измерения
$\eta _i $ известен. Если мы ищем минимум эмпирической ошибки, то из теоремы представления следует, что если мы выберем $\mathcal{H}_{(AA^{\ast })^{-1}}$ в качестве 
функционального пространства, в котором мы ищем искомый минимум
\begin{equation}
\label{eq:EmpiricalError}
\mathcal{E} (h) = \frac{1}{2} \sum_{i=1}^N \left( \eta _i - h({\bm x}_i ) \right)^2 + \gamma \Vert h \Vert ^2_{\mathcal{H}_{(AA^{\ast })^{-1}}} ,
\end{equation}
то этот минимум $h^{\ast }$ может быть записан как
\begin{displaymath}
h^{\ast } ({\bm x} ) = \sum_{i=1}^N  \alpha _i (f \ast f)({\bm x} - {\bm x}_i ) ,
\end{displaymath}
где $\alpha _i \in \mathbb{R} $ — подходящие коэффициенты. Параметр регуляризации Тихонова $\gamma $ (см., например, \cite{Hansen}) обычно выбирается таким образом, чтобы амплитуды 
коэффициентов $\alpha _i $ оставались
в пределах некоторых заранее выбранных границ, в то же время обеспечивая как можно меньшую эмпирическую ошибку. Заметим, что существуют и другие подходы, которые не используют явно параметр регуляризации.
Из приведённой формулы непосредственно следует выражение для восстанавливающей функции $t^{\ast }$, которая минимизирует эмпирическую ошибку (\ref{eq:EmpiricalError})
\begin{equation}
\label{eq:dwellTime}
t^{\ast } ({\bm x} ) = \sum_{i=1}^N  \alpha _i f ({\bm x} - {\bm x}_i ) .
\end{equation}
Формулируя более абстрактно, задачу поточечного решения уравнения $At = h$ проще решить, находя подходящие пробные функции в пространстве результатов измерений, чем используя
базис в пространстве восстанавливающих функций. Для пробных функций, построенных из ядра оператора $AA^{\ast }$, соответствующие восстанавливающие функции очевидны, при условии что мы знаем ядра
операторов $A$ и $A^{\ast }$. Стоит отметить, что в зависимости от точных свойств оператора $A$, гильбертово пространство $\mathcal{H}_{(AA^{\ast })^{-1}}$ может быть значительно меньше, чем
$L^2 \left( \Omega , \lambda \right)$. Это наблюдение согласуется с эвристическим ожиданием, что, за исключением шума, функции измерений, наблюдаемые на практике, должны быть
довольно гладкими.

Определяя $N$-мерные векторы ${\bm \alpha } = ( \alpha _i ) $, ${\bm \eta } = ( \eta _i ) $ и матрицу $ K = (k ({\bm x}_i , {\bm x}_j) )$, эмпирический функционал (\ref{eq:EmpiricalError})
может быть представлен как
\begin{equation}
\label{eq:FiniteDimMeritFunction}
\mathcal{E} ({\bm \alpha }) = \frac{1}{2} \langle {\bm \alpha } , K^{T} K {\bm \alpha } \rangle - \langle {\bm \alpha } , K {\bm \eta } \rangle + \frac{1}{2} \langle {\bm \eta },  {\bm \eta } \rangle + \gamma
 \langle {\bm \alpha } , K {\bm \alpha } \rangle ,
\end{equation}
что является конечномерным аналогом целевой функции (\ref{eq:QuadMeritFunction}). На практике целевая функция (\ref{eq:FiniteDimMeritFunction}) используется вместе с 
граничными условиями и ограничениями,
которые отражают физику задачи. Например, для расчётов карты времени экспозиции в фигурировании ионным пучком, время экспозиции $t^{\ast } ({\bm x} )$ должно быть
положительным\footnote{В пространстве функций, определённых линейными суммами вида (\ref{eq:dwellTime}), карты времени экспозиции образуют положительный конус.}, т.е. $t^{\ast } ({\bm x} ) \geq 0$.

\subsection{Дискретный случай и радиальные базисные функции}

Сделанные до сих пор замечания применимы к случаю, когда восстанавливающие функции непрерывны. Однако в некоторых экспериментальных установках машин для фигурирования ионным пучком
ионные пучки позиционируются в дискретном наборе локаций, так что математически карта травления $t({\bm x})$ представляет собой конечную сумму дельта-мер
\begin{displaymath}
t_{d} ({\bm x}) = \sum_{j=1}^M \alpha _j \delta _{{\bm y}_j } ({\bm x}) ,
\end{displaymath}
на некотором конечном множестве точек ${\bm y}_j$, $j=1,...,M$, по аналогии с ситуацией, рассмотренной в разделе \ref{ss:DiscreteCase}, см. соотношение (\ref{eq:discreteT}). Отметим
также, что точки ${\bm y}_j$ не обязательно совпадают с точками измерений ${\bm x}_i$, в которых производятся измерения.

С точки зрения дискретизации измерительного сигнала и в свете формулы (\ref{eq:discreteH}) представляется удобным аппроксимировать результаты измерений
$\eta _i $ с помощью взвешенных сумм функций $f ({\bm x} - {\bm y}_j )$, $j=1,...,M$, которые можно понимать как радиальные базисные функции
\begin{equation}
\label{eq:RBS}
h ({\bm x} ) = \sum_{j=1}^M  \alpha _j f ({\bm x} - {\bm y}_j ) .
\end{equation}
Затем времена экспозиции могут быть вычислены путём минимизации эмпирического функционала ошибки типа (\ref{eq:EmpiricalError}) при дополнительном ограничении, а именно,
что травления должны быть положительными. Это условие легко выполнить в дискретном случае, так как оно идентично условию $\alpha _i \geq 0$.
Отметим, что описанный здесь подход очень близок к методу, предложенному в \cite{Chernyshev} (см. также \cite{ChernyshevPhD}).

\section{ Фильтрация результатов измерений с помощью разложения на собственные функции} 
\label{s:Filtering}

Мы видели, что подход на основе гильбертовых пространств с воспроизводящим ядром (RKHS) очень хорошо подходит для поточечного вычисления восстанавливающих функций. 
Однако этот подход нельзя использовать
без предварительной фильтрации данных, потому что, во-первых, он не различает вклады времени экспозиции на основе их эффективности восстановления, как это имеет место
в функционально-аналитическом подходе, отражённом в уравнении (\ref{eq:pseudoinverse}) раздела \ref{ss:InfiniteDimLinEq}.
Во-вторых, мы ожидаем, что шум измерений в основном происходит от мелкомасштабных (высокочастотных) особенностей, которые, как мы видели в 
уравнении (\ref{eq:AsymptoticEigenvalue}), соответствуют малым собственным значениям оператора $A^{\ast }A$ (или, что
идентично, $AA^{\ast }$). Таким образом, при использовании подхода, основанного на использовании гильбертовых пространств с воспроизводящими ядрами, шум должен быть 
удалён до подгонки данных методами, подобными представленным в \cite{Petrakov}.
Вопрос эффективности можно решить косвенно, ограничивая величину второго члена в эмпирическом функционале ошибки (\ref{eq:EmpiricalError}).

Как уже упоминалось в разделе \ref{s:Practice}, в практических приложениях мы можем использовать разложение результатов измерений
$h({\bm x})$ по собственным функциям $AA^{\ast }$ (или левым сингулярным векторам $A$) $h_n$ и усечение на выбранном номере собственного вектора $N$ как процедуру
фильтрации данных измерений, по крайней мере, пока порядок усечения $N$ остаётся в разумных пределах. Из-за плохих свойств сходимости численной
схемы, основанной на вычислении скалярных произведений
$\langle h_n , h \rangle $, мы подгоняем усечённые суммы собственных функций $h_n$ поточечно к рассматриваемым результатам измерений, минимизируя эмпирический функционал.

В непрерывном случае мы имеем
\begin{equation}
\label{eq:EmpiricalErrorMeas}
\mathcal{E} (h) = \frac{1}{2} \sum_{i=1}^N \left( \eta _i - \sum_{n=1}^{n_{Tr} } c_n h_n ({\bm x}_i ) \right)^2 +
\gamma \Vert \sum_{n=1}^{n_{Tr} } c_n h_n \Vert ^2_{\mathcal{H}_{(AA^{\ast })^{-1}}} .
\end{equation}
По коэффициентам $c_n$, которые минимизируют приведённый выше эмпирический функционал, мы строим отфильтрованные измерения $\sum_{n=1}^{n_{Tr} } c_n h_n ({\bm x} )$.
Отсюда мы можем двигаться в двух направлениях. В первом случае полученные коэффициенты используются для вычисления восстанавливающей функции
\begin{equation}
\label{eq:contDwellTime}
t^{\star } ({\bm x} ) = \sum_{n=1}^{n_{Tr} } \frac{c_n}{\sqrt{\lambda _n }} t_n ({\bm x} )  .
\end{equation}
Эта формула была использована в разделе \ref{s:Practice} для вычисления времён экспозиции для различных порядков усечения.
На тот же эмпирический функционал мы можем наложить дополнительные ограничения, такие как условие положительности, чтобы гарантировать, что мы вычисляем положительные восстанавливающие функции.
Во втором случае альтернативные процедуры, например, такие, которые более адекватно отражают свойства машин, выполняющих коррекцию поверхности, могут быть
применены к отфильтрованным данным измерений для получения восстанавливающих функций.


Если мы приближённо знаем масштаб $l_{noise}$ структур, которые мы хотим удалить как шум измерений, мы можем легко оценить
порядок усечения $n_{Tr}$ для функций $t_n $ и $h_n$, по крайней мере в гауссовом случае. А именно, собственное значение, соответствующее этому порядку, должно удовлетворять
соотношению
\begin{equation}
\label{eq:LengthScale}
\frac{2\pi }{l_{noise}} \approx \frac{1}{\sigma } \sqrt{ -\ln \lambda _{n_{Tr}} } .
\end{equation}
Эта оценка может быть использована в качестве ориентира для практических вычислений карт травления.

Наконец, следует сделать замечание относительно практических аспектов вычисления карты травления или восстанавливающей функции. Конкретный вид процедуры, используемой для этой цели,
сильно зависит от рассматриваемой задачи. Рассмотрим в качестве примера процесс коррекции поверхности зеркала посредством ионно-лучевого травления. По мере развития производства зеркал
становится необходимым контролировать несколько параметров управления, которые характеризуют работу зеркала в оптической схеме, например, помимо стреднеквадратичного отклонения поверхности 
зеркала от идеальной часто необходимо учитывать влияние этих отклонений на качество волнового фронта всей оптической схемы. Кроме того, другие величины,
являющиеся производными вычисленной карты травления, такие как время обработки зеркала, должны поддерживаться в разумных пределах. Также в промышленном производственном 
процессе обычно не ставят целью достижение наилучших возможных значений параметров управления, что лишь неоправданно увеличило бы стоимость производства, а стремятся удерживать их ниже
некоторых предопределённых уровней допуска. Следовательно,
процедура, основанная на условной оптимизации, с использованием таких методов, как достижение цели \cite{Gembicki} или некоторых их вариантов, может быть выбрана в качестве подходящего подхода 
для задачи вычисления карты травления.

\section{Выводы}
\label{s:conclusion}

Данная статья посвящена разработке математического аппарата, подходящего для определения карт травления ионным пучком по заданным профилям коррекции поверхности. 
Мы рассмотрели два подхода: более абстрактный, использующий инструменты функционального анализа, и более практический,
применяющий гильбертовы пространства с воспроизводящими ядрами и радиальные базисные функции. Второй подход легко реализовать на практике, так как он позволяет вычислять
карты травления непосредственно из данных измерений, но он не даёт глубокого понимания сути задачи, которая рассматривается задача аппроксимации
данных. Этот подход может быть вполне пригоден в определённых обстоятельствах, но он требует для данных измерений выполнения некоторых предварительных условий. Результаты
измерений должны быть отфильтрованы до аппроксимации, и более того, мы требуем, чтобы поведение данных измерений на границах области измерений было более или менее регулярным. 
Также ожидается, что соответствие между аппроксимацией и исходными данными будет лучше внутри области, чем вблизи границ.

В первом подходе мы формулируем проблему как
бесконечномерное линейное уравнение (\ref{eq:Ax=b}), что позволяет нам рассматривать целый класс обратных задач с приложениями в различных областях. Оказывается, что, поскольку
в задаче возникают ядерные операторы, наша проблема не слишком далека от стандартных конечномерных линейных уравнений и, по крайней мере, формальное решение
может быть построено с использованием псевдообратных операторов. Проблема устойчивости решения становится ещё более важной, чем в конечномерном случае, потому что
сингулярные значения $A$ будут сгущаться в нуле, который является единственно возможной предельной точкой спектра. 
Мы также видим, что сингулярные значения оператора, связанного с нашим бесконечномерным уравнением, могут
интерпретироваться как эффективность реализации соответствующих сингулярных векторов. В разделе \ref{s:EigsAAstar} мы показываем, что в случае, когда размер инструмента мал
по сравнению с областью измерений, мы можем дать грубую оценку структуры сингулярных значений и затем описать сингулярные векторы внутри
области измерений. Действительно, они оказываются близки к собственным функциям бесконечно глубокой квантовой ямы. Стоит отметить, что в случае гауссовой
функции инструмента собственные функции оператора $A^{\ast } A$ являются аналитическими функциями, которые представляют собой своего рода локализацию функций косинуса и синуса
на области измерений и стремятся к нулю как экспонента квадрата расстояния от границы вне области.

Поскольку сингулярные векторы $A$ ведут себя как функции косинуса и синуса внутри области измерений, мы можем связать с ними длину волны, которая, по крайней мере
в гауссовом случае, уменьшается с уменьшением сингулярного значения, см. формулу (\ref{eq:LengthScale}). Таким образом, усечение количества сингулярных векторов действует,
грубо говоря, как усечение в пространстве Фурье и, следовательно, может использоваться для фильтрации шума. Как было отмечено ранее, это также исключает компоненты карт травления 
с низкой эффективностью реализации.

Не существует общего ответа на вопрос, какой из двух вышеупомянутых подходов следует использовать для практического решения конкретных задач.
В зависимости от деталей, один из обоих подходов или их комбинация могут давать лучшие результаты. Мы лишь упомянем, что в установке травления ионными пучками, 
типа описанной в \cite{Chernyshev}, могут использоваться комбинации обоих подходов: усечённое разложение данных измерений
по сингулярным векторам, которое представляет собой эффективный фильтр измерений, даёт ограничение оптимизации для аппроксимации данных измерений с помощью радиальных базисных
функций (\ref{eq:RBS}).

Отметим, что функционально-аналитический аппарат, представленный здесь, может быть легко расширен за пределы трансляционно инвариантных ядер, что становится необходимым
при обработке криволинейных поверхностей. Кроме того, мы можем также рассмотреть ситуацию, когда заданная поверхность последовательно полируется несколькими инструментами с
различными функциями инструмента, и мы вычисляем карты травления для каждого из них, см. \ref{app:SeveralBeams}. Дальнейший интересный аспект — это роль формы области измерений.
Как мы заметили, правые сингулярные векторы, описывающие время экспозиции, близки к собственным функциям бесконечно глубокой квантовой ямы. Как мы знаем, в зависимости от
формы ямы, собственные значения и собственные функции оператора Гамильтона могут демонстрировать сильно различающееся поведение. Не лишен интереса вопрос , в какой степени 
такие явления могут оказываться значимыми для практического вычисления карт травления и других связанных с этим проблем.

\bigskip

Автор благодарит Влада Желиговского и Алексея Чернышева за полезные замечания и обсуждения. Он также благодарен Самриддхи Шанкар Раю, Дарио Винченци, 
Такеси Мацумото, Константину Ханину и Леониду Гораю за помощь и поддержку. Работа выполнена при поддержке гранта РНФ № 21-72-30029-П.

\appendix

\section{Некоторые сведения из функционального анализа}
\label{app:FuncAna}

Чтобы сделать статью более доступной для широкой аудитории физиков и инженеров и для удобства читателей, мало знакомых с 
функциональным анализом, мы поясняем здесь некоторые понятия и результаты, используемые в настоящей статье.

\subsection{Некорректность и регулярность}

Понятие некорректности уравнения вместе с сопровождающими граничными условиями восходит к работам Адамара в начале 
двадцатого века. Уравнение является корректно поставленным, когда выполнены три условия: (i) решение существует, (ii) это решение единственно, 
(iii) решение устойчиво по отношению к изменению
начальных условий, см. \cite{ThikhonovArsenin}. Заметим, что с математической точки зрения существование решения также означает, что, например,
функция, его описывающая, принадлежит некоторому априори выбранному классу функций с известными свойствами гладкости, интегрируемости или ограниченности.
Регулярность решений одинаково важна для их единственности, как стало ясно на примере решений двумерных уравнений Эйлера, которые 
единственны для гладких начальных условий и допускают бесконечно много решений для начальных условий без достаточной регулярности \cite{Szekelyhidi}.
Однако, наиболее важным для практических приложений считается вопрос устойчивости решений по отношению к начальным условиям, см. \cite{BenningBurger}.
Грубо говоря, это условие означает, что малые изменения начальных условий должны вызывать малые изменения в соответствующих решениях.

Случай преобразования Вейерштрасса, упомянутый в разделе \ref{eq:Preliminary}, может служить хорошим примером проблем регулярности, связанных с обращением уравнений типа 
свертки (\ref{eq:convolution1}). Для функции $t(x^{\prime })$ её преобразование Вейерштрасса $h(x)$ определяется как 
\begin{displaymath}
h(x) = \frac{1}{\sqrt{4\pi \sigma ^2 } } \int_{\mathbb{R} }  e^{-\frac{(x - x^{\prime })^2 }{4 \sigma ^2} }  t(x^{\prime }) \, dx^{\prime } .
\end{displaymath}
Как показано в \cite{BrychkovPrudnikov}, см. формулу (311) русского издания, функция $t(x^{\prime })$ действительно может быть вычислена по $h(x)$, при условии 
что $t(x^{\prime })$ трактуется как обобщённая функция, потому что решение имеет смысл только в рамках обобщённых функций. Этот результат означает, что 
уравнение свёртки (\ref{eq:convolution1}) может быть решено содержательным образом только при значительно ослабленных предположениях о регулярности решения. 

\subsection{Операторы и их свойства}

Пусть $\mathcal{H}$ — гильбертово пространство со скалярным произведением $\langle \rangle _{\mathcal{H} } $. Для наших целей достаточно рассматривать пространства 
$L^2 \left( \mathbb{R}^n , \lambda \right)$ или $L^2 \left( \Omega , \lambda \right)$ квадратично-интегрируемых функций на $\mathbb{R}^n $ или на ограниченной области $\Omega $.
Заметим, что пространства $L^2 \left( \mathbb{R}^n , \lambda \right)$ и $L^2 \left( \Omega , \lambda \right)$ являются частными случаями более общих банаховых пространств 
$L^p \left( \mathbb{R}^n , \lambda \right)$ или $L^p \left( \Omega , \lambda \right)$.
Здесь $L^p \left( \mathbb{R}^n , \lambda \right)$ — это пространство интегрируемых по мере Лебега $\lambda $ функций, $L^p$-норма которых, определённая как 
\begin{displaymath}
\Vert t \Vert _{L^p \left( \mathbb{R}^n , \lambda \right)} = \left( \int_{\mathbb{R}^n } \vert t({\bm x} ) \vert^p \, d^n {\bm x} \right)^{\frac{1}{p} } ,
\end{displaymath}
конечна. Пространство $L^p \left( \Omega , \lambda \right)$ определяется аналогично.

Пусть задан линейный оператор $A: \mathcal{H} \longrightarrow \mathcal{H}^{\prime } $. Область значений, которые $A$ может принимать, обозначается как $R(A) \subset  \mathcal{H}^{\prime } $, 
а ядро $A$ обозначается как $N(A) \subset \mathcal{H}$.
Оператор $A$ называется ограниченным, если существует конечная константа $0 \leq C < \infty $, зависящая от 
$A$, такая что для всех $t \in \mathcal{H}$ выполняется 
\begin{displaymath}
\Vert A t \Vert _{\mathcal{H}^{\prime } } \leq C \Vert t \Vert _{\mathcal{H} } .
\end{displaymath}
Наименьшее вещественное число, для которого выполняется приведённое неравенство, называется операторной нормой $A$ и обозначается $\Vert A \Vert _{\infty }$. Из приведённой формулы следует, что
композиция двух ограниченных операторов является ограниченным оператором.

Чтобы показать ограниченность оператора $A$ в разделе \ref{ss:InfiniteDimLinEq}, мы использовали неравенство Юнга для свёртки (см. \cite{BrascampLieb}), которое утверждает, что для функций 
$f \in L^p \left( \mathbb{R}^n , \lambda \right)$ и 
$t \in L^q \left( \mathbb{R}^n , \lambda \right)$ выполняется соотношение
\begin{displaymath}
\Vert f \ast t \Vert _{L^r \left( \mathbb{R}^n , \lambda \right)} \leq \Vert f \Vert _{L^p \left( \mathbb{R}^n , \lambda \right)}
\Vert t \Vert _{L^q \left( \mathbb{R}^n , \lambda \right)} , 
\end{displaymath}
где $p,q,r$ — вещественные числа $1 \leq p,q,r \leq \infty $, такие что 
\begin{displaymath}
\frac{1}{p} + \frac{1}{q} = \frac{1}{r} + 1 .
\end{displaymath}
Полагая $p=1$ и $q = r = 2$, мы получаем неравенство 
\begin{displaymath}
\Vert A t \Vert _{L^2 \left( \Omega , \lambda \right)} \leq \Vert f \ast t \Vert _{L^2 \left( \mathbb{R}^n , \lambda \right)} \leq \Vert f \Vert _{L^1 \left( \mathbb{R}^n , \lambda \right)}
\Vert t \Vert _{L^2 \left( \mathbb{R}^n , \lambda \right)} , 
\end{displaymath}
из которого следует, что $A$ является ограниченным оператором с операторной нормой $\Vert A \Vert _{\infty }$, равной $\Vert f \Vert _{L^1 \left( \mathbb{R}^n , \lambda \right)}$. 

Сопряжённый оператор к $A$ определяется как единственный оператор $A^{\ast }: \mathcal{H}^{\prime } \longrightarrow \mathcal{H}$ (см. параграф 12.9 в \cite{Rudin}), удовлетворяющий следующему соотношению
\begin{displaymath}
\langle A s , t \rangle _{\mathcal{H} } = \langle s , A^{\ast } t \rangle _{\mathcal{H} } , \qquad \forall s,t \in \mathcal{H} , 
\end{displaymath}
для которого также выполняется $\Vert A^{\ast } \Vert _{\infty } = \Vert A \Vert _{\infty }$.
Ограниченный оператор $B$ называется самосопряжённым, если $B^{\ast } = B$. Заметим, что это простое определение применимо только к ограниченным операторам, тогда как для неограниченных операторов должны выполняться дополнительные требования.
Из вышесказанного следует, что оператор $A^{\ast } A$ является ограниченным и самосопряжённым.

Компактность оператора $A$, определённого в разделе \ref{ss:InfiniteDimLinEq}, может быть показана с использованием критериев компактности в \cite{KantAkil}, глава IX, для чего
достаточно доказать, что $A$ отображает единичный шар $B_n$ в $L^2 \left( \mathbb{R}^n , \lambda \right)$ в относительно компактное множество. Поскольку $A$ является ограниченным оператором, он отображает 
единичный шар в ограниченное множество. Затем мы можем использовать теорему 1.2 из \S 1, ibid., чтобы доказать относительную компактность $A B_n$, показав, что усреднённые значения 
$At$ равномерно сходятся к $At$ для $t \in B_n$, когда функция влияния инструмента непрерывна (см. \S 2, ibid.), что имеет место для всех примеров, кроме Примера 3. Функция влияния инструмента из 
Примера 3 является просто частным случаем функций усреднения по значению, для которых относительная компактность показана в доказательстве упомянутой теоремы 1.2. Из компактности $A$ следует,
что $A^{\ast } A$ также является компактным оператором.

\subsection{Спектр оператора и спектральное разложение}

Если $B$ — компактный самосопряжённый оператор в гильбертовом пространстве $\mathcal{H}$, то существует ортонормированный базис из собственных векторов $s_n$ оператора $B$. Множество соответствующих собственных значений ограничено и
стремится к нулю с ростом порядка собственного значения $\lim _{n \to \infty} \lambda _n = 0$, см., например, \cite{Simon}. Тогда, учитывая кратность собственных значений, $B$ может быть записан как 
\begin{displaymath}
Bs = \sum_{n=1}^{\infty } \lambda _n \langle s_n , s \rangle _{\mathcal{H} } s_n .  
\end{displaymath}
Это спектральная теорема в её простейшей форме, которая выполняется для компактных операторов.

Оператор $B$ называется неотрицательно определённым, если  $\langle Bt , t \rangle _{\mathcal{H} } \geq 0 $ для всех $t \in \mathcal{H}$. Собственные значения таких операторов неотрицательны. Пример 
такого оператора — $A^{\ast } A$, поскольку 
\begin{displaymath}
\langle A^{\ast} At , t \rangle _{\mathcal{H} } = \langle At , At \rangle _{\mathcal{H} }  = \Vert At \Vert_{\mathcal{H} }^2 \geq 0 .
\end{displaymath}

Для общих компактных операторов мы имеем разложение в терминах его левых и правых сингулярных векторов $t_n $ и $h_n $, определённых посредством (\ref{eq:singVectors})
\begin{displaymath}
A t = \sum_{n=1}^{\infty } \sqrt{\lambda _n } \langle t_n , t \rangle _{\mathcal{H} } h_n ,
\end{displaymath}
которое иногда называют разложением по сингулярным значениям оператора $A$. Здесь $\lambda _n $ обозначают собственные значения оператора $A^{\ast } A$.

\subsection{Оценки операторов}

Помимо операторной нормы $\Vert \cdot \Vert _{\infty }$, введённой выше, существуют различные другие методы оценки операторов, которые применяются для выделения определённых 
подклассов операторов, см. \cite{Simon}. 
Норма Гильберта--Шмидта $\Vert \cdot \Vert _{HS} $, определённая в разделе \ref{ss:PropertiesAAStar}, возникает при изучении интегральных операторов и интегральных уравнений.
Другие типы оценок операторов могут быть построены с использованием понятия следа оператора. Для заданного ортонормированного базиса $s_n$, след оператора определяется как 
\begin{displaymath}
Tr (B) = \sum_{n=1}^{\infty } \langle B s_n  , s_n \rangle _{\mathcal{H} } ,
\end{displaymath}
при условии, что он не зависит от выбора базиса. Для $A^{\ast } A$ след может быть вычислен как 
\begin{displaymath}
Tr (A^{\ast } A) = \sum_{n=1}^{\infty } \lambda _n  ,
\end{displaymath}
где $\lambda _n $ — собственные значения $A^{\ast } A$. Для интегральных операторов с непрерывным и положительным ядром след $Tr (A^{\ast } A)$ может быть вычислен 
с помощью теоремы Мерсера, см. теорему 2.12 в \cite{Simon} и предложение 5.6.9 в \cite{BrianDavies}, что даёт 
\begin{displaymath}
Tr (A^{\ast } A) = \sum_{n=1}^{\infty } \lambda _n  = \int_{\mathbb{R}^n } k ( {\bm x} , {\bm x} )  \, d^n {\bm x} ,
\end{displaymath}
и приводит к формуле (\ref{eq:traceNorm}) в разделе \ref{ss:PropertiesAAStar}.

Норма Шаттена--фон Неймана для оператора $B$ определяется как $\Vert B \Vert _2 = \sqrt{Tr (B^{\ast } B)} $. Норма Шаттена--фон Неймана для $A^{\ast } A$ может быть определена
через соотношение
\begin{displaymath}
\Vert A^{\ast } A \Vert _2^2 = Tr (A^{\ast } A A^{\ast } A ) = \sum_{n=1}^{\infty } \langle A^{\ast } A t_n  , A^{\ast } A t_n \rangle _{\mathcal{H} } = \sum_{n=1}^{\infty } \lambda ^2_n .
\end{displaymath}
В теореме 2.11 в \cite{Simon} показано, что норма Шаттена--фон Неймана равна норме Гильберта--Шмидта, откуда следует оценка (\ref{ineq:HS}).

\bigskip

\section{Задача вычисления карт травления при использовании нескольких пучков}
\label{app:SeveralBeams}

В данной работе рассматривается вопрос вычисления карты травления для одного пучка. Если в заданной карте корреции профиля поверхности $h({\bm x})$ 
присутствуют структуры сразу на нескольких масштабах, то кажется рациональным, в первую очередь для уменьшения общего времени травления 
использовать для коррекции сразу несколько пучков с размерами, соответствующими масштабам вышеупомянутых структур. 

Рассмотрим $N$ функций пучка $f^{(i)}({\bm x})$. Обозначим искомые карты травления, или, другими словами, карты времен экспозиции как $t^{(i)}({\bm x})$.
Тогда изменение поверхности в результате травления равняется $\sum_{i=1}^{N} f^{(i)} \ast t^{(i)}$. Пусть целевая функция задана через среднеквадратичное 
отклонение изменения поверхности в результате травления от измеренного профиля поверхности
\begin{displaymath}
\Phi (\{ t^{(i)} \} ) =\frac{1}{2} \langle \sum_{i=1}^{N} f^{(i)} \ast t^{(i)} - h, \sum_{i=1}^{N} f^{(i)} \ast t^{(i)} - h \rangle _{L^2 \left( \Omega , \lambda \right)} .
\end{displaymath}
Из этого выражения сразу видно, что если определить операторы 
\begin{displaymath}
f^{(i)} \ast t = A_{i}:L^2 \left( \mathbb{R}^n , \lambda \right) \longrightarrow L^2 \left( \Omega , \lambda \right) ,
\end{displaymath}
то целевую функцию можно записать в виде
\begin{displaymath}
\Phi (\{ t^{(i)} \} ) =\frac{1}{2} \sum_{i=1}^{N} \sum_{j=1}^{N} \langle A_i^{\ast } A_j t^{(j)} ,  t^{(i)} \rangle _{L^2 \left( \Omega , \lambda \right)}) - 
\sum_{i=1}^{N} \langle A_i^{\ast } h , t^{(i)}  \rangle _{L^2 \left( \Omega , \lambda \right)} +
\frac{1}{2} \langle h ,  h \rangle _{L^2 \left( \Omega , \lambda \right)} ,
\end{displaymath}
где $A_i^{\ast }$ --- оператор сопряженный к $A_i$. Тогда условие первого порядка для минимизации целевой функции принимает вид 
следующей системы, состоящей из $N$ уравнений 
\begin{displaymath}
\sum_{j=1}^{N}  A_i^{\ast } A_j t^{(j)}  = A_i^{\ast } h ,
\end{displaymath}
где $i = 1,\ldots ,N$. Таким образом, используя формализм, представленный в разделе \ref{s:FuncAna},
мы приходим к необходимости рассмотрения собственных значений и собственных функций операторов $A_i^{\ast } A_j$.

Рассмотрим для примера две гауссовы функции пучка с ширинами $\sigma _1 $ и $\sigma _2 $. Тогда ядра 
операторов $A_i^{\ast } A_j$  задаются выражениями
\begin{eqnarray*}
& & k_{i,j} (x^{\prime } , x^{\prime \prime }) = \int_{-L}^L \frac{1}{\sqrt{2\pi \sigma _i^2 }} e^{-\frac{(x-x^{\prime })^2 }{2 \sigma _i^2} }
\frac{1}{\sqrt{2\pi \sigma _j^2 }} e^{-\frac{(x-x^{\prime \prime })^2 }{2 \sigma _j^2} } \, dx =
\\
& & \frac{1}{\sqrt{2 \pi (\sigma _i^2 + \sigma _j^2)} } \, 
e^{-\frac{(x^{\prime } - x^{\prime \prime } )^2 }{2(\sigma _i^2 + \sigma _j^2)} } \, \frac{1}{2} \, \biggl[ 
\mathrm{erf} \frac{(x^{\prime } + L) \sigma _j^2 + (x^{\prime \prime } + L) \sigma_i^2 }{ \sigma _i \sigma _j \sqrt{2\sigma _i^2 + 2 \sigma _j^2} }  -  
\\
& & \mathrm{erf} \frac{(x^{\prime } - L) \sigma _j^2 + (x^{\prime \prime } - L) \sigma_i^2 }{ \sigma _i \sigma _j \sqrt{2\sigma _i^2 + 2 \sigma _j^2} } 
\biggr]
\end{eqnarray*}
Заметим, что вышеописанный подход легко обобщить также и на случай непрерывно изменяющейся функции пучка.


\begin{thebibliography}{99}
\bibitem{JPhysA}
W.~Pauls 2025 \textit{J. Phys. A: Math. Theor.} {\bf 58} 475201
\bibitem{Patent}
W.~Pauls, C.~Petri, S.~Vauth, U.~Kubon 2020 DE102019209575A1
\bibitem{WM87} 
S.R.~Wilson, J.R.~McNeil, 1987 \textit{Proc. SPIE} {\bf 818} 320
\bibitem{WRM88} 
S.R.~Wilson, D.W.~Reicher, J.R.~McNeil 1988 \textit{Proc. SPIE} {\bf 966} 74
\bibitem{1} 
T.W.~Drueding, S.C.~Fawcett, S.R.~Wilson, T.G.~Bifano 1995 \textit{Opt. Eng.} {\bf 34} 3565
\bibitem{2} 
T.~Frantz, T.~H\"ansel 2008 \textit{Optical fabrication and testing} (Rochester, New York, United States) p 21
\bibitem{6}
M.~Zeuner, S.~Kiontke 2012 \textit{Optik \& Photonik} {\bf 7} 56
\bibitem{5} 
M.~Xu, Y.~Dai, X.~Xie, L.~Zhou, W.~Liao 2015 \textit{Appl. Optic} {\bf 54} 8055
\bibitem{7} 
C.~Jiao, S.~Li, , X.~Xie 2009 \textit{Appl. Optic} {\bf 48} 4090
\bibitem{8} 
T.~Wang, L.~Huang, H.~Kang , H.~Choi, D.W.~Kim, K.~Tayabaly, M.~Idir 2020 \textit{Sci. Rep.} {\bf 10} 4090
\bibitem{9}
C.L.~Carnal, C.M.~Egert, K.W.~Hylton 1992 \textit{Proc. SPIE} {\bf 1752} 54
\bibitem{10} 
T.W.~Drueding, T.G.~Bifano, S.C.~Fawcett 1995 \textit{Precis. Eng.} {\bf 17} 10
\bibitem{HirschWidd} 
I.I.~Hirschmann, D.V.~Widder 2005 \textit{The Convolution Transform} (Dover Publications, Mineola, New York)
\bibitem{Nair}
M.T.~Nair, Ill-posedness of backward heat conduction problem, 
National Seminar on “Mathematical Modelling” at PSGR Krishnammal College for Women, Coimbatore on August 6, 2004
\bibitem{BrychkovPrudnikov}
Yu.A.~Brychkov, A.P.~Prudnikov 1989 \textit{Integral Transforms of Generalized Functions} (Gordon and Breach Science Publishers)
\bibitem{BenisraelGreville}
A.~Ben-Israel, T.N.E.~Greville 2003 \textit{Generalized Inverses} (Springer)
\bibitem{CheverdaKostin}
V.A.~Cheverda, V.I.~Kostin 1995 \textit{J. Inv. Ill-Posed Problems} {\bf 3} 131
\bibitem{BlinMoes}
S.~Blinnikov , R.~Moessner 1998 \textit{Astronomy and Astrophysics. Supplement Series} {\bf 130} 193
\bibitem{Vazquez}
J.L.~V\'azquez 2017 \textit{Complex Variables and Elliptic Equations} {\bf 63} 1
\bibitem{Bertram}
S.~Bertram 1940 \textit{Proceedings of the I.R.E.} {\bf 9} 418
\bibitem{KantAkil}
L.V.~Kantorovitch, G.P.~Akilov 2014 \textit{Functional Analysis} (Elsevier)
\bibitem{BrianDavies}
E.~Brian Davies 2007 \textit{Linear Operators and Their Spectra} (Cambridge University Press)
\bibitem{BrascampLieb}
H.J.~Brascamp, E.H.~Lieb 1976 \textit{Advances in Mathematics} {\bf 20} 151
\bibitem{BenningBurger}
M.~Benning, M.~Burger 2018 \textit{Modern regularization methods for inverse problems} (Cambridge University Press)
\bibitem{WhittakerWatson}
E.~T.~Whittaker, G.~N.~Watson  2012 \textit{A Course of Modern Analysis} (Watchmaker Publishing)
\bibitem{AbramowitzStegun}
M.~Abramowitz, I.A.~Stegun 1983 \textit{Handbook of Mathematical Functions With Formulas, Graphs, and Mathematical Tables} (Dover Publications)
\bibitem{NgGeller}
E.W.~Ng, M.~Geller 1969 \textit{Journal of Research of the National Bureau of Standards, Section B: Mathematical Sciences} {\bf 73B} 1
\bibitem{Gutzwiller}
M.C.~Gutzwiller 1990  \textit{Chaos in Classical and Quantum Mechanics} (Springer, New York)
\bibitem{Garcia}
A.G.~Garc\'{\i}a 2014 \textit{Operator Theory} (Springer, Basel) 1
\bibitem{Zeidler}
E.~Zeidler 1995 \textit{Applied Functional Analysis} (Springer, New York)
\bibitem{BahChemD}
H.~Bahouri, J.-Y.~Chemin, R.~Danchin 2011 \textit{Fourier Analysis and Nonlinear Partial Differential Equations} (Springer)
\bibitem{Hansen}
P.C.~Hansen 1998  \textit{Rank-Deficient and Discrete Ill-Posed Problems} (SIAM)
\bibitem{Chernyshev}
A.~Chernyshev, N.~Chkhalo, I.~Malyshev, M.~Mikhailenko, A.~Pestov, R.~Pleshkov, R.~Smertin, M.~Svechnikov, M.~Toropov 2021  \textit{Precision Engineering} {\bf 69} 29
\bibitem{ChernyshevPhD}
А.К.~Чернышев 2025 \textit{Развитие ионно-пучковых методов формирования восокоточных поверхностей для рентгеновской оптики} (кандидатская диссертация, ИФМ РАН)
\bibitem{Petrakov}
E.V.~Petrakov, E.I.~Glushkov, A.K.~Chernyshev, N.I.~Chkhalo 2024 \textit{Optical Engineering} {\bf 63} 114104
\bibitem{Gembicki}
F.W.~Gembicki 1974 \textit{Vector Optimization for Control with Performance and Parameter Sensitivity Indices} (PhD thesis, Case Western Reserve University)
\bibitem{ThikhonovArsenin}
A. N.~Tikhonov,  V. Y.~Arsenin 1977 \textit{Solutions of ill-Posed Problems} (Winston, New York)
\bibitem{Szekelyhidi}
L.~Sz\'ekelyhidi Jr. 2012 \textit{Journ\'{e}es \'{E}quations aux Deriv\'{e}es Partielles} {\bf 10} 491
\bibitem{Rudin}
W.~Rudin 1991 \textit{Functional Analysis} (International Series in Pure and Applied Mathematics. Vol. 8 (Second ed.). New York, NY)
\bibitem{Simon}
B.~Simon 2005 \textit{Trace Ideals and Their Applications} (AMS)
\end{thebibliography}
\end{document}